\documentstyle[aps,prl,twocolumn,eqsecnum,floats,axodraw]{revtex}
\def\Journal#1#2#3#4{{#1} {\bf #2}, #3 (#4)}

\def\NPB{Nucl. Phys. {\bf B}$\!$}

\def\PLB{Phys. Lett. B}

\def\PRD{Phys. Rev. D}

\def\JHEP{JHEP}
 
\def\KeyWord#1{$\backslash$\IfColor{$\!\!$\textRed{#1}\textBlack}{#1}$\!\!$}

\def\ordere{$\cal{O}(\epsilon)$} 
\begin{document}
\wideabs{
\title{DREDed Anomaly Mediation}

\author{Ed Boyda$^{1}$, Hitoshi Murayama$^{1,2}$, and Aaron 
Pierce$^{1,2}$.} 
\address{
1. Department of Physics, 
University of California, 
Berkeley, CA~~94720, USA;\\ 
2. Theory Group, 
Lawrence Berkeley National Laboratory, 
Berkeley, CA~~94720, USA}
\date{\today}
\maketitle
\begin{abstract}

We offer a guide to dimensional reduction (DRED) in theories with
anomaly mediated supersymmetry breaking.  Evanescent operators
proportional to $\epsilon$ arise in the bare Lagrangian when it is
reduced from $d=4$ to $d=4-2\epsilon$ dimensions.  In the course of a
detailed diagrammatic calculation, we show that inclusion of these
operators is crucial.  The evanescent operators conspire to
drive the supersymmetry-breaking parameters along anomaly-mediation
trajectories across heavy particle thresholds, guaranteeing the
ultraviolet insensitivity.
\end{abstract}
}

\narrowtext

\setcounter{footnote}{0}
\setcounter{page}{1}
\setcounter{section}{0}
\setcounter{subsection}{0}
\setcounter{subsubsection}{0}

\section{Introduction}

Anomaly mediation is a remarkably predictive framework for
supersymmetry breaking in which the breaking of scale invariance
mediates between hidden and visible
sectors~\cite{RandallSundrum,anomalymed}.  Since the soft
supersymmetry-breaking parameters are determined by the breaking of
scale invariance, they can be written in terms of beta functions and
anomalous dimensions in relations which hold at all energies.  An
immediate consequence is that supersymmetry-breaking terms are
completely insensitive to physics in the ultraviolet.  Anomalous
dimensions and beta functions, which depend only on degrees of freedom
excitable at a given energy, completely specify the soft parameters at
that energy.  This property makes anomaly mediation an attractive
solution to the supersymmetric flavor problem.  The low-energy
spectrum of soft masses and couplings is independent of the physics
that explains flavor in the ultraviolet.\footnote{This property has
lead to the well known issue of tachyonic sleptons.  People have taken
various approaches towards solving this
problem~\cite{RandallSundrum,nondecoupling,KSS} which jeopardize the
ultraviolet insensitivity.  However, it was shown recently that the UV
insensitivity can be preserved while solving the problem of tachyonic
sleptons \cite{viableUV}.}

On the other hand, Regularization by Dimensional REDuction (DRED)
\cite{DRED} is often the preferred regulator for supersymmetric field
theories.  As with ordinary dimensional regularization (DREG), DRED is
simpler computationally than Pauli-Villars or other cutoff methods. 
DRED is also superior to DREG in that it preserves supersymmetry: In
DREG when we analytically continue the dimension of space-time away
from $d=4$, the spinor algebra changes, creating a mismatch between
fermionic and bosonic degrees of freedom.  DRED avoids this problem by
compactifying from $d=4$ to $d=4-2\epsilon$ dimensions and making the
fields independent of the extra $2\epsilon$ dimensions.  The spinor
algebra doesn't change, so the regulated theory is still
supersymmetric.

In this paper we explore the subtleties of DRED in theories with
anomaly mediated supersymmetry breaking.  We point out is that it is
not correct to just add anomaly-mediated supersymmetry breaking to the
Lagrangian if DRED is used.  Since most calculations in the literature
are done this way, our result raises a warning flag.  In retrospect,
it is not surprising why it is so.  In the case of the chiral anomaly,
one does not add the chiral anomaly as an additional term to the
Lagrangian.  When the theory is properly regularized, the chiral
anomaly is the outcome rather than a part of the input Lagrangian.
Similarly, the anomaly-mediated supersymmetry breaking must be the
outcome of the Lagrangian rather than the additional terms in the bare
Lagrangian.  We show that the most important consequence of
compactifying to $4-2\epsilon$ dimensions is the introduction of
evanescent operators, proportional to $\epsilon$, into the bare
Lagrangian.  These operators prove to be of first importance in
diagrammatic anomaly-mediation calculations.  Proper inclusion of
these operators yields a DRED-based formalism suitable for anomaly
mediation calculations.  In addition we discuss the implications of
DRED's failure to regulate infrared divergences, which follows because
the dimension of space-time is necessarily $d<4$ in DRED.

As a showcase for our DRED-based anomaly mediation formalism, we
perform an explicit diagrammatic calculation that shows the
ultraviolet insensitivity of anomaly mediation.  Although the
appearance of supersymmetry-breaking parameters and the decoupling of
flavor physics have been well-understood through the spurion formalism
(see \cite{nondecoupling} for the most comprehensive review of
anomaly mediation using the spurion formalism), the phenomena have not
been investigated in a diagrammatic framework.  The spurion analysis
fixes the $A$-terms to be
\begin{equation}\label{eqn:atermform}
A_{ijk}= - m_{3/2}\lambda_{ijk}(\gamma_{i} +\gamma_{j} +\gamma_{k}),
\label{eq:A}
\end{equation}
while scalar masses are given by
\begin{equation}\label{eqn:scalarform}
\tilde{m}^{2}_{i}=\frac{1}{2} |m_{3/2}|^2 \dot{\gamma}_{i}.
\label{eq:m2}
\end{equation}
Here, $m_{3/2}$ is the gravitino mass, $\lambda_{ijk}$ is the
superpotential Yukawa coupling, $\gamma_{i} \equiv -\frac{1}{2} \mu
\frac{d}{d \mu} \log Z_{i}$ is the anomalous dimension of the $i^{th}$
superfield, and $\dot{\gamma}_{i} \equiv \mu \frac{d}{d \mu}
\gamma_{i}$.  To fix signs, these terms appear in the Lagrangian as
\begin{equation}
{\cal L} \ni - \tilde m_i^2 \tilde Q_i^*\tilde Q_i -A_{ijk}\tilde
Q_i\tilde Q_j\tilde Q_k + \textrm{h.c.},
\end{equation}
with scalar fields $\tilde Q_i$.  It is highly non-trivial that the
forms in Equations~(\ref{eq:A}) and (\ref{eq:m2}) indeed are invariant
under the renormalization-group evolution, which was checked
explicitly in \cite{JJ}.  We apply our DRED calculation to see in
detail how various diagrams conspire to set the soft parameters on
their anomaly mediated trajectories across the massive particle
thresholds.  In particular, loops containing evanescent $\epsilon$
operators produce the soft terms above the threshold of flavor
physics, and additional evanescent operators combine with the flavor
fields to decouple the flavor sector below threshold.  We find that
when calculating with DRED, it is inconsistent to simply insert the
soft terms of Equations~(\ref{eqn:atermform})
and~(\ref{eqn:scalarform}) into the Lagrangian while neglecting the
evanescent operators.

In Section \ref{sec:regularize} we review some established results of
anomaly mediated supersymmetry breaking.  In section \ref{sec:DREDful}, we present a puzzle that makes clear the need to develop a consistent framework for using DRED with anomaly mediation.  In Section \ref{sec:deriveL} we focus on developing this framework, deriving the dimensionally reduced bare Lagrangian.  In Section \ref{sec:deriveSUSYbreaking}, we utilize this Lagrangian to discuss the origin of
Equations~(\ref{eqn:atermform}) and (\ref{eqn:scalarform}).  Having
established a framework for using DRED with anomaly mediation, we demonstrate
its use through explicit diagrammatic calculations which confirm the UV
insensitivity of anomaly mediation.  In section \ref{sec:moral}, we take a
moment to recapitulate, and emphasize the basic message of our derivation of
the anomaly mediated DRED-based formalism.  Section
\ref{sec:Model} defines the models used in our diagrammatic
calculations.  In section \ref{sec:ATerms} we compute the $A$-terms, a
short one-loop calculation.  In section \ref{sec:ScalarMasses} we
discuss the substantially more complicated case of the scalar
masses, which is a two-loop calculation.

\section{Anomaly Mediation and Holomorphic Regularization} \label{sec:regularize}

In this section, we provide a brief review of established results in anomaly mediated supersymmetry breaking.  We discuss the origin of the anomaly mediated contributions.  We also review the spurion analysis for regularization schemes that use an explicit cut-off.  This discussion will provide a useful foil for the DRED scheme which we later employ.
   
In anomaly mediated models of supersymmetry breaking 
\cite{RandallSundrum,anomalymed}, the sole source of
supersymmetry breaking resides in the chiral compensator field in the
supergravity Lagrangian: $\langle \Phi \rangle = 1 + m_{3/2}
\theta^2$.  We now review the origin of this field.  Supergravity is not scale-invariant because it has an
explicit mass scale: the Planck scale.  However, it is possible to
reformulate the theory as conformal supergravity by compensating for the
non-invariance of the Lagrangian under Super-Weyl transformations by a fictitious transformation of the
chiral compensator field $\Phi$.  

The supersymmetry breaking that arises when chiral compensator takes on its vacuum expectation value will always be present.  However, in general, $M_{pl}$ suppressed operators coupling the ``observable sector'' to the ``hidden sector'' often dominate over these contributions.  Nevertheless, the chiral compensator can dominate the supersymmetry breaking effects, for example, if the
``observable'' sector (including the Supersymmetric Standard Model)
and the ``hidden sector'' (responsible for supersymmetry breaking)
reside on different branes in extra dimensions \cite{RandallSundrum} or if
the dynamics of the hidden sector is nearly super-conformal to suppress
direct couplings between the hidden and observable fields in the
K\"ahler potential \cite{Luty:2001jh}.  In these cases, the only communication of
supersymmetry breaking effects from the hidden to the observable
sector occurs through the supergravity multiplet, and hence the
auxiliary component of the chiral compensator field.  Since the
coupling of the chiral compensator is completely fixed by the
(fictitious) super-Weyl invariance, the consequent supersymmetry
breaking terms in the observable sector are highly constrained. 
This case, where the couplings between the observable and hidden
sector are suppressed and the form of the SUSY breaking is highly
restricted, is known generically as anomaly mediation, and it is the
case which we discuss here.

If the observable sector does not have explicit mass scales, the
Lagrangian is scale-invariant at the classical level.  Then the
coupling of the chiral compensator can be completely eliminated from
the Lagrangian by appropriate redefinition of the fields.  However,
the scale invariance is broken at the quantum level because of the
need to regulate the theory.  This leads to residual couplings of the
chiral compensator to the observable fields.  When the classical
invariance of the Lagrangian is broken at the quantum level leading to
physical effects, this is generically called an ``anomaly.''  This
explains the name ``anomaly mediated supersymmetry breaking.''

The implementation previously discussed in the literature uses an explicit cutoff
scale $\Lambda$.  Because of the imposed super-Weyl invariance,
the cutoff parameter $\Lambda$ only appears in the combination
$\Lambda \Phi$.  Such a cutoff is possible using Pauli--Villars
regulators, finite $N=2$ theories \cite{Arkani-Hamed:2000mj}, or higher
derivative regularization \cite{KapLouis}.  Any of these
methods preserve manifest supersymmetry, and the cutoff is a
holomorphic parameter: The cutoff can be viewed as the lowest
component of a chiral superfield.  We refer to all these schemes generically as
``holomorphic regularization.''  If a holomorphic regularization scheme is
used, independent of the details of the regularization method, we can
derive their consequences on the supersymmetry breaking effects in the
observable fields as follows.  

The matter kinetic terms receive wave function renormalization
\begin{equation}
\label{eqn:renKahler}
    \int d^{4}\theta {\cal Z}_{i} Q_{i}^{*} Q_{i},
\end{equation}
Here, ${\cal Z}_{i}$ is the superfield extension of the wave-function renormalization, $Z_{i}$, following the formalism developed in \cite{nimaetal,Giudice}. ${\cal Z}_{i}$ depends on the cutoff
\begin{equation}\label{eqn:logZ1}
    \log {\cal Z}_{i}(\mu) = \sum_{k=1}^{\infty} C_{k}
    \log^{k} \frac{(\Lambda \Phi)(\Lambda\Phi)^{\dagger}}{\mu^{2}}.
\end{equation}
Here, $C_{k}$ are functions of dimensionless coupling constants. 
Expanding the logarithms in $\theta$,
\begin{eqnarray}
    \lefteqn{
    \log {\cal Z}_{i}(\mu) = \log Z_{i}(\mu) } \nonumber \\
    & & + (\theta^{2} m_{3/2} + \bar{\theta}^{2} m_{3/2})
    \sum_{k=1}^{\infty} k C_{k} 
    \log^{k-1} \frac{(\Lambda \Phi)(\Lambda\Phi)^{\dagger}}{\mu^{2}}
    \nonumber \\
    & & 
    + \theta^{2} \bar{\theta}^{2} m_{3/2}^{2}
    \sum_{k=1}^{\infty} k(k-1) C_{k}
    \log^{k-2} \frac{(\Lambda \Phi)(\Lambda\Phi)^{\dagger}}{\mu^{2}} 
    \nonumber \\
    &=& \log Z_i(\mu) 
    + (\theta^{2} m_{3/2} + \bar{\theta}^{2} m_{3/2}) \gamma_{i}
    - \frac{1}{2} \theta^{2} \bar{\theta}^{2} m_{3/2}^{2}
	\dot{\gamma}_{i}. \nonumber \\
    &=& \log Z_{i}(\mu)
    - (\theta^{2} A_{i} + \bar{\theta}^{2} A_{i}^{*})
    - \theta^{2} \bar{\theta}^{2} m_{i}^{2}.
\label{eqn:PVderive}
\end{eqnarray}
Here, $\gamma = - \frac{1}{2} \mu \frac{d}{d\mu} \log Z$ and
$\dot{\gamma} = \mu \frac{d}{d\mu} \gamma$.   The identification of the soft terms (the last line of Equation~(\ref{eqn:PVderive})), follows from rescaling the fields in Equation~(\ref{eqn:renKahler}) by $Q_{i} \to \frac{Q_{i}}{1+ \gamma m_{3/2} \theta^{2}}$. Once we note that $A_{ijk} = A_{i} + A_{j} + A_{k}$, this 
leads to the predictions in Equations~(\ref{eq:A},\ref{eq:m2}).  As an aside, we note that both
$\gamma(\mu)$ and $\dot{\gamma}(\mu)$ must be finite once re-expressed
in terms of the running coupling constants at the scale $\mu$.    

The gauge coupling constant is given in terms of the bare coupling
$1/g_{0}^{2}$ and the running effects in the Wilsonian effective
Lagrangian as
\begin{equation}
    \int d^{2} \theta \left( \frac{1}{g_{0}^{2}} + 
    \frac{b_{0}}{8\pi^{2}} \log \frac{\Lambda \Phi}{\mu} 
    - \sum_{f} \frac{T_{f}}{8\pi^{2}} \log \left. {\cal Z}_{i} 
    \right|_{\bar{\theta}=0} \right) 
    W_{\alpha} W^{\alpha}.
\end{equation}
By expanding the logarithms to ${\cal O}(\theta^{2})$, we 
find the prediction for the holomorphic gaugino mass
\begin{equation}
    m_{\lambda}(\mu) = -\frac{g_{0}^{2}}{8\pi^{2}} \left(b_{0} - \sum_{f} T_{f} 
    \gamma_{f}(\mu)\right) m_{3/2}.
\end{equation}
Going to the canonical normalization of the gaugino changes the above expression to 
\cite{HS}
\begin{equation}
    m_{\lambda} = -\frac{g^{2}(\mu)}{8\pi^{2}} \frac{b_{0} - \sum_{f} T_{f} 
    \gamma_{f}}{1-\frac{g^{2}(\mu)}{8\pi^{2}} C_{A}} m_{3/2}
    = -\frac{\beta(g)}{2g^{2}} m_{3/2}.
\end{equation}

To complete our review of established anomaly mediated results, 
we reemphasize that anomaly mediation possesses the property of
ultraviolet insensitivity, namely that the effects of heavy particles
completely decouple from the supersymmetry breaking effects in the
low-energy theory.  With a holomorphic regularization, this is quite
easy to see. 
Instead of logarithms dependent on $\mu$, as in
Equation~(\ref{eqn:logZ1}), loop effects of a heavy particle cutoff at
its mass $M$ and so appear with the logarithms
\begin{equation}
    \log 
    \frac{(\Lambda\Phi)(\Lambda\Phi)^{\dagger}}{(M\Phi)(M\Phi)^{\dagger}}
    = \log \frac{\Lambda^{2}}{M^{2}}.
\label{eqn:proofHoloReg}
\end{equation}
The point here is that the super-Weyl invariance makes the mass $M$
appear only in the combination $M\Phi$ which precisely cancels the
corresponding $\Phi$ dependence of the cutoff.  Therefore there are no
supersymmetry breaking effects from heavy particles in the low-energy 
theory.  We will now attempt to understand this ultraviolet insensitivity explicitly in the DRED formalism as well.

\section{DRED-ful UV sensitivity?}\label{sec:DREDful}

In this section, we will outline a naive DRED calculation.  We will
find that simply adding the anomaly mediated soft terms of
Equations~(\ref{eq:A}) and (\ref{eq:m2}) to our Lagrangian by hand and
then calculating using DRED leads to inconsistencies.  In particular,
we are unable to recover the well-established result of UV
insensitivity.  In this section, we demonstrate the problem using the
technique of Arkani-Hamed, Giudice, Luty and Rattazzi \cite{nimaetal}
which ``analytically continues'' parameters in the Lagrangian to the
full superspace to incorporate the effects of soft supersymmetry
breaking.  We will do explicit diagrammatic calculations in later
sections to further illuminate this problem.

Consider a simple Yukawa model
\begin{equation}
    {\cal W} = h \tau X_{1} X_{2} + M X_{1} Y_{1} + M X_{2} Y_{2},
\end{equation}
where $\tau$ is a light field and $X_{i}$, $Y_{i}$ heavy.  The 
massive fields have tree-level supersymmetry breaking because the 
chiral compensator appears in the superpotential as $M\Phi$:
\begin{equation}
    {\cal L}_{\it soft} = -M m_{3/2} (\tilde{X}_{1} \tilde{Y}_{1} + 
    \tilde{X}_{2} \tilde{Y}_{2}) + \textrm{h.c.}
    \label{eq:tree}
\end{equation}
In addition, there are anomaly mediated effects according 
to the general formula of Equations ~(\ref{eq:A},\ref{eq:m2}),
\begin{eqnarray}
    {\cal L}_{\it soft} &=& - 3 \frac{(h^{*}h)^{2}}{(4\pi)^{4}}
    m_{3/2} (\tilde{\tau}^{*} \tau + \tilde{X}_{1}^{*} \tilde{X}_{1} +
    \tilde{X}_{2}^{*} \tilde{X}_{2}) \nonumber \\ & &- 3
    \frac{h^{*}h}{(4\pi)^{2}} m_{3/2} h \, \tilde{\tau} \tilde{X}_{1}
    \tilde{X}_{2} + \textrm{h.c.}.  \label{eq:anomaly}
\end{eqnarray}
The question of ultraviolet insensitivity is whether the scalar mass for the $\tilde{\tau}$ shown in Equation~(\ref{eq:anomaly}) is precisely canceled 
by the threshold effects from $X$, $Y$ loops.

As we will describe in detail in Section~\ref{sec:ScalarMasses}, the
loops of $X$ and $Y$ precisely cancel $m_{\tilde{\tau}}^{2}$, if all
integrals are done in four-dimensions, paying careful attention to keep
all integrals finite.  However, we can also understand this
computation rather simply using the language of the spurions.  
First, we compute the $Z$-factor
for $\tau$ at $Q^{2} \gg M^{2}$.  It is given by
\begin{equation}
    \log Z_{\tau} (Q) = \frac{h^{*}h}{(4\pi)^{2}} \log 
    \frac{|\Lambda|^{2}}{Q^{2}}
    - \frac{(h^{*}h)^{2}}{(4\pi)^{4}} \frac{3}{2} 
    \log^{2} \frac{|\Lambda|^{2}}{Q^{2}}.
\end{equation}
Now, we incorporate supersymmetry-breaking effects by substituting
$\Lambda \rightarrow \Lambda \Phi$.  Performing this replacement, and
inserting the vacuum expectation value for the chiral compensator,
$\langle \Phi \rangle =1 + m_{3/2} \theta^{2}$, we obtain the anomaly
mediated pieces shown in Equation ~(\ref{eq:anomaly}).  Now we
integrate between the scale $Q$ and $M$ and find the low-energy
theory below $M$.  The additional contribution to $Z_{\tau}$ is
\begin{equation}
    \Delta \log Z_{\tau} = \frac{h^{*}h}{(4\pi)^{2}} \log 
    \frac{Q^{2}}{M^{2}}
    - \frac{(h^{*}h)^{2}}{(4\pi)^{4}} \frac{3}{2} \log^{2} 
    \frac{Q^{2}}{M^{2}}.
\end{equation}

Using this expression, we can isolate the supersymmetry-breaking
effects in the threshold correction.  

One effect arises from taking $M \rightarrow M\Phi$ in the last term,
which gives $\Delta m_{\tilde{\tau}}^{2} = +3
\frac{(h^{*}h)^{2}}{(4\pi)^{4}} m_{3/2}^{2}$.  This corresponds to the
sum of all two-loop diagrams in Figures~\ref{fig:h4scalars} and
\ref{fig:h4fermion} with $-Mm_{3/2} \tilde{X_{i}} \tilde{Y_{i}}$ mass
insertions.  The other source of SUSY breaking is the $A$-term.  Its
effects can be obtained by the replacement $h \rightarrow h (1 - 3
\frac{h^{2}}{(4\pi)^{2}} m_{3/2} \theta^{2})$ together with $M
\rightarrow M\Phi$ in the first term (and a similar replacement for
$h^{*}$).  The contribution to $\Delta m_{\tilde{\tau}}^{2}$ is $-6
\frac{(h^{*}h)^{2}}{(4\pi)^{4}} m_{3/2}^{2}$.  This corresponds to the
one-loop diagram,{\sl Graph~7-1}, that contains one $A$-term and one
$Mm_{3/2}$ mass insertion.  Adding the threshold corrections to the anomaly
mediated piece $+3 \frac{(h^{*}h)^{2}}{(4\pi)^{4}} m_{3/2}^{2}$, we
find a complete cancellation.  This cancellation demonstrates the UV
insensitivity.

Now we perform the same calculations, using regularization by
Dimensional REDuction (DRED), and we do not find the complete
cancellation.  The threshold correction can again be read off from the
$Z$-factor
\begin{eqnarray}
    \lefteqn{
    \Delta \log Z_{\tau}}
    \nonumber \\
    &=& \frac{h^{*}h}{(4\pi)^{2}} 
    (M^{-2\epsilon} - Q^{-2\epsilon}) \frac{1}{\epsilon}
    - \frac{(h^{*}h)^{2}}{(4\pi)^{4}} \frac{3}{2} 
    (M^{-4\epsilon} - Q^{-4\epsilon}) \frac{1}{\epsilon^{2}}.
    \nonumber \\
    \label{eq:DeltaZ}
\end{eqnarray}
Suppose we do the calculation in the same spirit as in the case with
the holomorphic regularization.  Then, we should again include
contributions from two sources: a cross term between an $A$-term and the
$Mm_{3/2}$ term, shown in {\sl Graph 7-1}, and the diagrams including only the
$Mm_{3/2}$ term.  The contribution from the $A$-term and $M\Phi$ in {\sl
Graph~7-1} can be found again by making the replacement $h \rightarrow
h (1 - 3 \frac{h^{2}}{(2\pi)^{2}} m_{3/2} \theta^{2})$ together with
$M \rightarrow M\Phi$ in the first term of Equation~(\ref{eq:DeltaZ}).
The result is the same as in the holomorphic regularization: $\Delta
m_{\tilde{\tau}}^{2} = -6 \frac{(h^{*}h)^{2}}{(4\pi)^{4}}
m_{3/2}^{2}$.  However the other contribution, from the
replacement $M \rightarrow M\Phi$ in the last term of
Equation~(\ref{eq:DeltaZ}), comes out differently.  Because
$(M^{2}\Phi\bar{\Phi})^{-2\epsilon} = (M^{2})^{-2\epsilon}(1 -
2\epsilon \theta^{2} m_{3/2} - 2\epsilon \bar{\theta}^{2} m_{3/2} +
4\epsilon^{2} \theta^{2} \bar{\theta}^{2} m_{3/2}^{2})$, we find
$\Delta m_{\tau}^{2} = +6 \frac{(h^{*}h)^{2}}{(4\pi)^{4}}
m_{3/2}^{2}$.
Summing this result with the contribution from the $A$-terms, we find
$\Delta m_{\tau}^{2}=0$.  We will explore in detail how this result,
which differs from holomorphic regularization result, arises in
section \ref{sec:ScalarMasses}.  For now, the important thing is to
realize that we have found an unexpected result.  We had hoped to find
a threshold correction, that when added to the anomaly mediated piece,
$+3 \frac{(h^{*}h)^{2}}{(4\pi)^{4}} m_{3/2}^{2}$, would yield a
complete cancellation.  Instead, we find that the ``threshold
correction'' itself vanishes.  Somehow we seem to have lost the
ultraviolet insensitivity!
\footnote{In fact, there is an additional piece that comes in at $h^{2}$ proportional to $\epsilon$.  The presence of this term does not change the effect that we have gotten an unexpected result.}

What we have seen here is that the naive addition of the anomaly mediated
supersymmetry breaking soft terms to a dimensionally-reduced theory leads to
incorrect results.  That is to say, putting the terms from Equation~(\ref{eq:A}) and (\ref{eq:m2}) in the Lagrangian by hand is {\it not} the correct prescription in DRED.  Note that most calculations in the literature are
done with this naive implementation.  We have to develop a consistent
formalism to implement anomaly mediated supersymmetry breaking within 
the DRED.  We proceed to do this in the following section.

Finally, we comment on the reason that things did not ``go wrong'' in
the holomorphic regularization scheme.  In that case, one has already
integrated out the fictitious Pauli-Villars fields at the cut-off
scale, yielding the anomaly-mediated soft terms of Equations
(\ref{eq:A}) and (\ref{eq:m2}) at the cut-off scale.  Therefore, in the
Pauli-Villars case, it is perfectly reasonable to treat the usual
anomaly mediated soft terms as a boundary condition at the cut-off
scale.  We will expand upon this point in section \ref{sec:moral}.

\section{Derivation of the Lagrangian in Dimensional Reduction}\label{sec:deriveL} 

In this section, we motivate the bare Lagrangians appropriate for use
with DRED regularization.  We look at both the case of a Yukawa theory
and a theory with gauge couplings, trusting that combining the two
provides no new wrinkles.  In each case, our procedure basically
consists of starting with a supersymmetric Lagrangian, and determining
how chiral compensators inject supersymmetry-breaking into the
Lagrangian.

By examining the Weyl scaling properties of the supergravity
fields~\cite{transforms}, we can determine where we must add chiral
compensator fields $\Phi$ to the supergravity Lagrangian to make it
super-Weyl invariant.  As noted above, we can then rescale fields so
that the chiral compensator appears only in front of dimensional
couplings.  This fixes how supersymmetry breaking enters the
Lagrangian since the breaking happens when $\Phi$ takes the vacuum expectation value
$\Phi = 1+m_{3/2}\theta^2$.

Here is how this works for a dimensionally reduced theory with Yukawa
couplings: In $4-2\epsilon$ dimensions, the Lagrangian, written in terms of bare chiral superfields looks like:
\begin{eqnarray}
\label{eqn:sugraguess}
\lefteqn{
{\cal L} = \int{ d^{4}\theta \;(\Phi \Phi^{\dagger})^{1-\epsilon}
Q_{i}^{\dagger} Q_{i}} 
} \nonumber \\
&-&  \left(\int{ d^{2} \theta \;
\Phi^{3-2\epsilon} (\lambda_{ijk,0} Q_{i} Q_{j} Q_{k}+ M_{ij,0} Q_{i}
Q_{j}) + {\mbox h.c}}\right).
\end{eqnarray}
Here, the $0$ subscript denotes a bare quantity. To recover canonical normalization, we rescale
\begin{equation}
Q_{i} \rightarrow \frac{Q_{i}}{\Phi^{1-\epsilon}}, 
\end{equation}
and then as promised, the chiral compensators only appear in front
of dimensionful couplings in the superpotential:
\begin{eqnarray}
\label{eqn:rescaleguess}
\lefteqn{
{\cal L} = \int{ d^{4}\theta \;Q_{i}^{\dagger} Q_{i}} 
} \nonumber \\ 
&-&
(\int{ d^{2}\theta \; \Phi^{\epsilon} \lambda_{ijk,0} Q_{i} Q_{j}
Q_{k}+ \Phi M_{ij,0} Q_{i} Q_{j} + {\mbox h.c}}).
\end{eqnarray}
The extra power of $\Phi^{\epsilon}$ can be thought of as arising from
the $\epsilon$ dimensionality of $\lambda_{ijk,0}$ which appears in $4-2
\epsilon$ dimensions.  Expanding in components, we find two sources of
supersymmetry breaking in the bare Lagrangian:
\begin{equation}\label{eqn:yukbreaking}
{\cal L}_{\it breaking} \ni -\epsilon m_{3/2}\lambda_{ijk,0} \tilde 
Q_{i}\tilde
Q_{j}\tilde Q_{k} - m_{3/2}M_{ij,0} \tilde Q_{i}\tilde
Q_{j}.
\end{equation}
The first term is one of the important evanescent operators which
produces anomaly mediated soft-terms to the low-energy effective
Lagrangian.

For the gauge theory we begin with the Lagrangian
\begin{equation}
{\cal L} \ni \frac{1}{4 g^{2}_{0}} \int d^{2} \theta \, W W + \frac{1}{4
g^{2}_{0}} \int d^{2} \bar\theta \, \bar W \bar W
\end{equation}
and dimensionally reduce it.  The $\Phi$ dependence can be fixed by
arguments of holomorphicity and dimensionality, in analogy with the
resulting $\Phi^{\epsilon}\lambda_{ijk,0}$ dependence found above.  Then
we should promote $\frac{1}{g^2_{0}}$ to a superfield gauge
coupling~\cite{Giudice,nimaetal},
\begin{equation}\label{eqn:S1}
\frac{1}{g^2_{0}}\to S = \frac{\Phi^{-2\epsilon}}{g^2_{0}},
\end{equation}
with which the Lagrangian becomes
\begin{equation}
{\cal L} \ni \frac{1}{4} \int d^{2} \theta \, S W W + \frac{1}{4} 
\int d^{2}\bar\theta \,  S^{\dagger}\bar W \bar W.
\end{equation}
We then would like to associate a real superfield, $R_{0}$, with the 
gauge coupling constant \cite{nimaetal}.  With the above Lagrangian, the superfield 
$R_{0}$, whose lowest component is $\frac{1}{g^{2}_{0}}$, is given by:
\begin{equation}\label{eqn:Rold}
R_{0} \equiv \frac{\Phi^{-2 \epsilon} + (\Phi^{\dagger})^{-2 
\epsilon}}{2g^{2}_{0}}.
\label{eq:R1}
\end{equation}  
However, this choice does not lead to the familiar prediction of the 
anomaly mediated supersymmetry breaking: $m^{2} = \frac{1}{2} 
\dot{\gamma} m_{3/2}^{2}$.  It differs at $O(\epsilon)$.  
We will work with a more convenient form 
that leads to the familiar prediction without $O(\epsilon)$ 
corrections.  Instead of Equation~(\ref{eqn:Rold}) we take
\begin{equation}
\label{eqn:Rnew}
R_{0} \equiv \frac{ (\Phi\Phi^{\dagger})^{-\epsilon}}{g^{2}_{0}}.
\label{eq:R2}
\end{equation}
The two expressions for $R_{0}$ differ only in $\theta^{2}
\overline{\theta}^{2}$ components, which does not lead to any physical
difference in the four-dimensional limit.  We prove this fact in the 
next section.   

Using Equation~(\ref{eqn:Rnew}) as the real gauge coupling superfield,
we can write the bare Lagrangian using the GMZ evanescent
operator~\cite{GMZ}.  Here the bare action is given by
\begin{equation} \label{eqn:GMZ}
\frac{1}{g^{2}_{0}}\int{d^{8} z} \, \frac{1}{\epsilon}(\Phi
\Phi^{\dagger})^{-\epsilon} g^{\mu \nu}_{\epsilon} \mbox{tr}\, (\Gamma_{\mu}
\Gamma_{\nu}).
\end{equation}
The metric tensor $g_{\epsilon}^{\mu\nu}$ runs only for the
compactified $2\epsilon$ dimensions, and $\Gamma_{\mu}$ is the gauge
connection defined by
\begin{equation}
\Gamma^{\mu} = \frac{1}{2} \sigma^{\mu}_{\alpha \dot{\alpha}} 
\overline{D}^{\dot{\alpha}} (e^{-V} D^{\alpha} e^{V}).
\end{equation}
This leads to a component Lagrangian that contains the following 
supersymmetry-breaking pieces:
\begin{eqnarray}
{\cal L}_{\it breaking} &\ni& \frac{1}{g^{2}_{0}} \left( \frac{1}{2} \epsilon m_{3/2}
\lambda \lambda +\frac{1}{2} \epsilon m_{3/2} \overline{\lambda}
\overline{\lambda} \right. \nonumber\\
& &\left. +\frac{\epsilon}{2}m_{3/2}^{2} g_{\epsilon}^{\mu\nu} A_{\mu} A_{\nu}\right)
\end{eqnarray}
Therefore, the supersymmetry-breaking effects are
a tree-level ${\cal O}(\epsilon)$ gaugino 
mass $m_{\lambda}=-\epsilon m_{3/2}$, and a tree-level $\epsilon$-scalar mass 
$m_{\epsilon}^{2} = \epsilon m_{3/2}^{2}$. 

For Abelian theories, we may also use
\begin{equation} \label{eqn:nonlocal}
\frac{1}{16 g^{2}_{0}}\int{d^{4} \theta} \, (\Phi
\Phi^{\dagger})^{-\epsilon}W^{\alpha} \frac{ {\cal D}^{2}}
{\Box} W_{\alpha} + {\mbox h.c.}
\end{equation}
to introduce the real superfield gauge coupling, Equation~(\ref{eq:R2}).
In this framework the $\epsilon$-scalar mass is replaced by a 
non-local modification of the gaugino propagator.  We find:
\begin{eqnarray}
\label{eqn:nonlocalL}
{\cal L}_{\it breaking} \ni \frac{1}{g^{2}_{0}} \left( \frac{1}{2} \epsilon m_{3/2}
\lambda \lambda +\frac{1}{2} \epsilon m_{3/2} \overline{\lambda}
\overline{\lambda}\right. \nonumber \\ 
\left.+\frac{1}{2} \epsilon^{2} m^{2}_{3/2}
\left(-i \frac{\lambda \sigma \cdot \partial 
\overline{\lambda}}{\Box}\right)\right).
\label{eqn:barenonlocal}
\end{eqnarray}
However, it is not clear how to interpret a non-local term in a bare
Lagrangian. Moreover an extension to non-Abelian theories is somewhat
opaque due to difficulties in making the expression containing 
$\frac{1}{\Box}$ gauge-covariant.  Nevertheless, it provides a useful 
cross-check to our calculations with the GMZ operator in an Abelian 
gauge theory.

\section{Derivation of the Soft supersymmetry Breaking Terms in 
DRED}\label{sec:deriveSUSYbreaking}

With bare Lagrangians in hand, we now go back and derive the anomaly
mediation formulas for the soft supersymmetry-breaking parameters
(Equations~(\ref{eqn:atermform}) and~(\ref{eqn:scalarform})) for DRED
regularization.  This discussion is to be compared with the known 
discussion for holomorphic regulators, reviewed in section 
\ref{sec:regularize}.

\subsection{Yukawa Theory}
In the Yukawa theory the bare Lagrangian is given by
Equation~(\ref{eqn:rescaleguess}).  For simplicity in this section we
drop the mass terms, so that 
\begin{equation}
{\cal L} = \int{ d^{4}\theta \;Q_{i}^{\dagger} Q_{i}} - 
 \left(\int{ d^{2}\theta \; \Phi^{\epsilon} \lambda_{ijk,0} Q_{i} Q_{j}
Q_{k} + {\mbox h.c}}\right).
\end{equation}
The important point is that $\Phi^{\epsilon} \lambda_{ijk,0}$ acts as an effective Yukawa
coupling constant.

We start by considering the wave-function re\-norm\-a\-li\-za\-tion%
\footnote{Note that in our notation, $Z^{-1}$ is the residue of the 
pole that one would find by calculating the two-point function.  That 
is to say, $Z$ would be the coefficient of 
the bare fields $Q Q^{\dagger}$ in the 1PI effective action.}  $Z$ that appears in the
effective Lagrangian.  Again, following the discussion of
\cite{nimaetal,Giudice}, we promote $Z$ to a superfield ${\cal Z}$,
and we expand in a power series of effective coupling constants
$\Phi^{\epsilon}\lambda_{ijk,0}$:
\begin{equation}\label{eqn:zexpand}
\log {\cal Z}(\mu)=\sum_{k=1}^{\infty} \frac{D_{k}}{\epsilon^{k}}\left(\frac{ 
\lambda_{ijk,0} \lambda_{ijk,0}^{*}(\Phi
\Phi^{\dagger})^{\epsilon}}{\mu^{2\epsilon}}\right)^{k} 
\end{equation} 
The coefficients $D_{k}$ are regular in the $\epsilon \rightarrow 0$
limit.  
Since the Yukawa coupling in $4-2 \epsilon$ dimensions is
dimensionful, it appears always with an appropriate factor of
$\Phi^{\epsilon}$.

Now we expand the chiral compensator $\Phi=1 + m_{3/2} \theta^{2}$, 
yielding the expression:
\begin{eqnarray}\label{eqn:DREDlogZ}
    \log {\cal Z} &=&
    \sum_{k=1}^{\infty} \frac{D_{k}}{\epsilon^{k}}\left(\frac{ 
    \lambda_{ijk,0} \lambda_{ijk,0}^{*}}{\mu^{2\epsilon}}\right)^{k} \nonumber \\
    & & + \sum_{k=1}^{\infty} \frac{(m_{3/2}\theta^{2}+ \mbox{h.c}) 
    k D_{k}}{\epsilon^{k-1}}\left(\frac{ 
    \lambda_{ijk,0} \lambda_{ijk,0}^{*}}{\mu^{2\epsilon}}\right)^{k} \nonumber \\
    & & + \sum_{k=1}^{\infty} \frac{m_{3/2}^{2} \theta^{2} \bar{\theta}^{2} k 
    ^{2} D_{k}}{\epsilon^{k-2}}\left(\frac{ 
    \lambda_{ijk,0} \lambda_{ijk,0}^{*}}{\mu^{2\epsilon}}\right)^{k}.
\end{eqnarray} 
Finally, we can write the expressions for $\gamma$ and 
$\dot{\gamma}$, by taking the appropriate derivatives of the first term in  
Equation~(\ref{eqn:DREDlogZ}).  We find
\begin{eqnarray}\label{eqn:DREDanomalous}
    & &\gamma=\sum_{k=1}^{\infty} \frac{k D_{k}}{\epsilon^{k-1}}\left(\frac{ 
    \lambda_{ijk,0} \lambda_{ijk,0}^{*}}{\mu^{2\epsilon}}\right)^{k} \\
    & &\dot{\gamma}=-2 \sum_{k=1}^{\infty} \frac{k^{2} 
    D_{k}}{\epsilon^{k-2}}\left(\frac{ 
    \lambda_{ijk,0} \lambda_{ijk,0}^{*}}{\mu^{2\epsilon}}\right)^{k}
\end{eqnarray}
Now using Equations~(\ref{eqn:DREDlogZ}, \ref{eqn:DREDanomalous}), and summing the 
contributions from the $i$, $j$, and $k$ particles, we find:
\begin{eqnarray}
\label{eqn:dimreg}
{\cal L}&=&\int{d^{4} \theta \left(1-\frac{\dot{\gamma_{i}}}{2} m_{3/2}^{2}
\theta^{2} \overline{\theta}^2 \right) Q^{\dagger}_{i} Q_{i}} \nonumber
\\& -& \int d^{2} \theta \lambda_{ijk} \Phi^{\epsilon} 
\left(1-
(\gamma_{i} + \gamma_{j} +\gamma_{k}) m_{3/2} \theta^{2}\right) Q_{i} Q_{j}
Q_{k} \nonumber \\ &+& \mbox{h.c.},
\end{eqnarray} 
where we distinguish the renormalized Yukawa coupling by
$\lambda_{ijk} \equiv \lambda_{ijk,0} Z_{i}^{-1/2} 
Z_{j}^{-1/2} Z_{k}^{-1/2}$.  The soft terms do indeed take the
form of Equation~(\ref{eqn:atermform}) and Equation~(\ref{eqn:scalarform}).
Notice, however, that an additional $O(\epsilon)$ supersymmetry-breaking Yukawa
coupling arises by expanding $\Phi^{\epsilon}$.  This is just the
tree-level evanescent operator from the bare Lagrangian as in
Equation~(\ref{eqn:yukbreaking}).  Our effective Lagrangian contains a
total $A$-term
\begin{equation}\label{eqn:fullaterm} 
A_{ijk}=-m_{3/2} \lambda_{ijk}(\gamma_{i} +\gamma_{j} 
+\gamma_{k}) +
\epsilon m_{3/2} \lambda_{ijk}.
\end{equation}


\subsection{Gauge Theory}

If we turn off the Yukawa theory but add gauge interactions, the
discussion proceeds analagously.  Instead of the effective Yukawa
coupling $\Phi^{\epsilon} \lambda_{ijk,0}$, the relevant expansion
parameter for ${\cal Z}$ is $g^{2}_{0} (\Phi \Phi^{\dagger})^{\epsilon}$.
This is clear from 
Equations~(\ref{eqn:GMZ},\ref{eqn:nonlocal}).  
Now we justify the
form of Equation~(\ref{eqn:Rnew}).  To do this, we need to show that
there is no physical consequence in switching from
Equation~(\ref{eqn:Rold}) to (\ref{eqn:Rnew}) in the four-dimesnional
limit.

Consider the following change in the real gauge-coupling superfield:
\begin{equation}
\label{eqn:Rtransform}
        R \rightarrow R + \theta^2 \bar{\theta}^2 \frac{\Delta^2}{g^2}
\end{equation}
Clearly this change will not affect one loop quantities such as
$A$-terms and the gaugino mass, as both of these depend solely on the
$\theta^{2}$ pieces of the Lagrangian.  We show now that the scalar
masses are also unaffected in the four dimensional limit as we pass
from Equation~(\ref{eqn:Rold}) to Equation~(\ref{eqn:Rnew}). 

The argument is simple.  (To keep our expressions uncluttered we work
with a single gauge coupling constant, but we have checked that the
argument can be generalized to multi-coupling theories.)  Generally,
under the transformation of $R$ in Equation~(\ref{eqn:Rtransform}),
the change in the mass-squared of a matter field $Q_{i}$ is
\begin{equation}
\label{eqn:theorem}
        m_i^2 \rightarrow m_i^2 + \frac{\gamma_{i}}{\epsilon} \Delta^2.
\end{equation}
We can see this as follows.
Starting from the expansion
\begin{equation}
\label{eqn:logZexp}
\log Z_i = \sum_{k=1}^\infty C_k g_0^{2k} \mu^{-2k\epsilon},
\end{equation}
we find
\begin{equation}
\gamma_{i} = \epsilon \sum_{k=1}^\infty k C_k g_0^{2k} \mu^{-2k\epsilon}.
\end{equation}
Now, the change in $R$ above is the same as the replacement
\begin{equation}
R^{-1} \rightarrow R^{-1} (1 - \theta^2 \bar{\theta}^2 \Delta^2) .
\end{equation}
Recall that to recover the scalar masses, we need the $\theta^{2} \bar{\theta^{2}}$ piece of $\log {\cal Z}$, which is found by replacing $g_0^{2}$ in Equation~(\ref{eqn:logZexp}) by $R^{-1}$.  So the change in $R^{-1}$ induces a change in $\theta^2 \bar{\theta}^2$ component of $\log {\cal Z}_i$ as 
given by making the replacement
\begin{equation}
g_0^{2} \rightarrow R^{-1} (1 - \theta^2 \bar{\theta}^2 \Delta^2)
\end{equation}
in Equation~(\ref{eqn:logZexp}).  Therefore the change in $m_i^2 = - 
\log {\cal Z}_i|_{\theta^2\bar{\theta}^2}$ is
\begin{equation}
\Delta m_{i}^{2}= - \sum_{k=1}^\infty C_k g_0^{2k} (-k \Delta^2) 
\mu^{-2k\epsilon} 
      = \frac{\gamma_{i}}{\epsilon} \Delta^2 .
\end{equation}
This proves the assertion of Equation~(\ref{eqn:theorem}).  Now notice that the difference between 
\begin{equation}
R_{1} = \frac{g_0^{-2} (\Phi^{-2\epsilon} + 
{{\Phi}^{\dagger}}^{-2\epsilon})}{2}
\end{equation}
and
\begin{equation}
R_{2} =  g_0^{-2} (\Phi {\Phi}^{\dagger})^{-\epsilon},
\end{equation}
is $\Delta^{2} \equiv R_{1} - R_{2}=-\epsilon^{2} m_{3/2}^{2}.$ Therefore, in this case, 
the change in the scalar masses is  only ${\cal O}(\epsilon)$ and 
does not affect the 4-dimensional limit.

Now that the choice $R_{2}^{-1} = g_0^{2} (\Phi 
\Phi^{\dagger})^{\epsilon}$ is justified, the derivation of the soft 
parameters follows the same path as in the Yukawa theory. 
Incidentally, our argument shows that we are performing a calculation
in the $\overline{DR}^{\prime}$ scheme.\cite{DRprime} We have
calculated the above-threshold case with a finite external momentum.
In particular, we can always take the value of this momentum to be
on-shell.  Then there is no additional change that depends
on the $\epsilon$-scalar mass in going from the $m^{2}(\mu)$ we have
calculated to the pole mass.
By definition, this is the $\overline{DR}^{\prime}$ scheme.  This is consistent with the 
comments found in \cite{nimaetal}.

We can also derive the gaugino mass following the same line, even 
though it was already discussed in \cite{anomalymed}.  The  
effective action is
\begin{equation} \label{eqn:GMZ2}
\int{d^{8} z} \, R(\mu) \frac{1}{\epsilon}(\Phi
\Phi^{\dagger})^{-\epsilon} g^{\mu \nu}_{\epsilon} tr (\Gamma_{\mu}
\Gamma_{\nu}).
\end{equation}
where the lowest order in $R(\mu)$ is the renormalized coupling
$g^{2}(\mu) \mu^{-2 \epsilon}$.  We know define a dimensionless
superfield, ${\cal F(\mu)}$, such that $g^{2}(\mu)= \left.  {\cal
F}^{-1}(\mu)\right|_{\theta=\bar{\theta}=0}$.   
The kinetic function is a function of the bare coupling $g_{0}$
together with the chiral compensator as
\begin{equation}
    {\cal F}(g_0^{2} \mu^{-2\epsilon} (\Phi\Phi^{\dagger})^{\epsilon}).
\end{equation}
Expanding the function ${\cal F}$, we find
\begin{eqnarray}
    {\cal F} (\mu) &=& \frac{1}{g^{2}(\mu)}
    + \left. {\cal F}'\right|_{\theta=\bar{\theta}=0} 
    g^{2}_{0} \mu^{-2\epsilon} \epsilon (\theta^{2} + 
    \bar{\theta}^{2}) m_{3/2}
    \nonumber \\
    & &+ \left({\cal F}' g^{2}_{0} \mu^{-2\epsilon} + 
    {\cal F}^{\prime\prime} g_0^{4} 
    \mu^{-4\epsilon}\right)_{\theta=\bar{\theta}=0} \epsilon^{2} 
    \bar{\theta}^{2}\theta^{2} m_{3/2}^{2}
\end{eqnarray}
Noting that
\begin{equation}
    \beta(g) = \mu \frac{d}{d\mu} \left. {\cal 
    F}^{-1}\right|_{\theta=\bar{\theta}=0}
    = -2\epsilon g^{2}_{0} \mu^{-2\epsilon}\left. \frac{-1}{{\cal F}^{2}}{\cal 
    F}'\right|_{\theta=\bar{\theta}=0} , 
\end{equation}
we find
\begin{equation}
    g^{2}_{0} \mu^{-2\epsilon} \epsilon 
    \left. {\cal F}' \right|_{\theta=\bar{\theta}=0}
    = \frac{\beta(g)}{2g^{4}(\mu)}.
\end{equation}
Furthermore, differentiating it on both sides,
\begin{eqnarray}
    & &-2\epsilon^{2} \left({\cal F}' g^{2}_{0} \mu^{-2\epsilon} + 
    {\cal F}^{\prime\prime} g^{4}_{0} 
    \mu^{-4\epsilon}\right)_{\theta=\bar{\theta}=0}
    = \mu \frac{d}{d\mu} \frac{\beta(g)}{2g^{4}(\mu)}
    \nonumber \\
    & & \qquad =  \frac{\dot{\beta}(g)}{2g^{4}(\mu)}
    - 2 \frac{\beta^{2}(g)}{2g^{6}(\mu)}.
\end{eqnarray}
Here, $\dot{\beta}(g) = \mu \frac{d}{d\mu} \beta (g)$.  Therefore,
\begin{eqnarray}
    {\cal F} (\mu) &=& \frac{1}{g^{2}(\mu)}
    + \frac{\beta(g)}{2 g^{4}(\mu)} (\theta^{2}+\bar{\theta}^{2})m_{3/2}
    \nonumber \\
    & & - \frac{1}{2} \mu\frac{d}{d\mu} \frac{\beta(g)}{2g^{4}(\mu)} 
    \theta^{2}\bar{\theta}^{2} m_{3/2}^{2}.
\end{eqnarray}
We therefore find the gaugino mass
\begin{equation}
    m_{\lambda} = - \frac{\beta(g)}{2 g^{2}(\mu)}m_{3/2},
\end{equation}
consistent with the derivation in \cite{anomalymed}.  We also find an 
all-order result for the epsilon scalar mass
\begin{equation}
    m_{\epsilon}^{2} = - \frac{1}{2} g^{2}(\mu) \mu\frac{d}{d\mu} 
    \frac{\beta(g)}{2g^{4}(\mu)}   
    m_{3/2}^{2}
\end{equation}
which had not been obtained in the literature.  It would be 
interesting to verify explicitly that this result is on the 
renormalization-group trajectory \`{a} la \cite{JJ}.

\section{Moral}\label{sec:moral}
The important moral to be taken away from the last three sections is
the following: in DRED anomaly mediated supersymmetry breaking effects
are to be {\it calculated}\/ from the bare Lagrangian, and cannot be
added to the Lagrangian by hand.  The basic mistake in the naive
calculation in Section~\ref{sec:DREDful} is that we added the
``anomaly mediated supersymmetry breaking'' to the Lagrangian by hand
and tried to demonstrate the UV insensitivity with this cobbled
together Lagrangian.  The reason why this is a mistake is clear from
the analogy to the chiral anomaly mentioned in the Introduction.  In a
regularized theory, the chiral anomaly comes out automatically from
the loop calculations.  One does not add the chiral anomaly as an
additional term to the Lagrangian.  In the same way, DRED is a
regularization, which leads automatically to the anomaly-mediated
supersymmetry breaking.  Therefore, instead of adding soft parameters
to the Lagrangian, we should perform a complete calculation starting
from the bare Lagrangian that contains a Yukawa coupling
$\lambda_0\Phi^{\epsilon}$ or a gauge coupling $g_0^{2}(\Phi
\Phi^{\dagger})^{\epsilon}$.  Then we should find that
contributions of heavy multiplets 
to the soft couplings vanish below the heavy mass threshold.  
We illustrate this UV insensitivity using DRED in our diagrammatic
calculation of Section \ref{sec:ScalarMasses}.

Finally, let us flesh-out this discussion by describing the proof of
ultraviolet insensitivity in anomaly mediation with the DRED framework.  This is the analogue
of Equation~(\ref{eqn:proofHoloReg}).  In general, the contributions
from heavy multiplets to the $Z$-factor have the dependence
$(\lambda^{*}\lambda)^{k} (M^{*} M)^{-k\epsilon}$.  The correct
inclusion of the chiral compensator then gives 
$(\lambda \Phi^{\epsilon} \lambda^{*} {\Phi^{\dagger}}^{\epsilon})^{k}
(M\Phi M^{*} \Phi^{\dagger})^{-k\epsilon} = (\lambda^{*}\lambda)^{k}
(M^{*} M)^{-k\epsilon}$, 
and no supersymmetry breaking effects remain.

Now we can analyze what went wrong in our example in section
\ref{sec:DREDful}.  Operationally, we made two errors in our
calculation.  First of all, we extended the Yukawa coupling
incorrectly.  Instead of extending it to get the $A$-term diagram
through the replacement $h \rightarrow h (1 - 3
\frac{h^{2}}{(4\pi)^{2}} m_{3/2} \theta^{2})$, we were meant to make
the replacement $h \rightarrow h \Phi^{\epsilon}$.  Moreover, we
neglected a one-loop ${\cal O}(\epsilon h^{2})$ piece that was present
in the high-energy theory.  In fact, in our attempt to compute the
threshold correction to $\log Z$, we ended up computing the entirety
of $\log Z$.  We were unable to separate the high-energy piece from
the threshold correction.

The above discussion seems to say that it is impossible to regard the
anomaly-mediated supersymmetry breaking as a boundary at the Planck
Scale.  Indeed, this appears to be true for DRED.
However, this is not impossible for other regularization schemes.  
We can take this view, for instance, if we use Pauli--Villars
regulators where the supersymmetry breaking $\Lambda m_{3/2}$ mass term for
the regulators is the source of all other supersymmetry breaking
effects.  Then we can play the following trick.  We add a pair of
correct- and wrong-statistics regulator fields without $\Lambda m_{3/2}$
mass term, which does not change the physics at all.  Then we
integrate out the original Pauli-Villars regulators with the $\Lambda
m_{3/2}$ mass term and a correct-statistics field without the $\Lambda
m_{3/2}$ mass term.
Integrating out this pair of fields will give us the soft
SUSY-breaking terms of Equations~(\ref{eq:A}) and (\ref{eq:m2}), where the
anomalous dimensions are to be evaluated at the cut-off scale.  The
left-over wrong-statistics massive field acts as the new
Pauli--Villars regulator while the supersymmetry breaking effects are
now in the Lagrangian.  This way, we obtain an entirely equivalent
theory with anomaly-mediated supersymmetry breaking in the bare
Lagrangian, regulated by the Pauli--Villars regulators that do not
have a $\Lambda m_{3/2}$ mass term.  On the other hand, DRED does not allow
us a similar trick because there is no ``regulator field.''  We need
to keep evanescent operators consistently in calculations.

In later sections, we will study the UV insensitivity with explicit
diagrammatic calculations.  The situation can be somewhat more subtle
in the presence of both light and heavy degrees of freedom, but
nonetheless we have demonstrated that the effects of heavy multiplets
completely disappear from the soft supersymmetry breaking parameters
below the heavy threshold once coupling constants are re-expressed in
terms of renormalized ones.

\section{Model Considered}\label{sec:Model}
We now define two simple toy models to fulfill the diagrammatic
computation promised in the previous sections.  Calculations using
these models will follow in sections \ref{sec:ATerms} and
\ref{sec:ScalarMasses}.

The first model contains only chiral superfields with minimal kinetic
terms and superpotential
\begin{equation}\label{eqn:w1}
{\cal W}_1= \lambda_{\tau,0}\tau LH +h_{0} \, \tau X_{1}
X_{2} + M \, X_{1} Y_{1} + M \, X_{2} Y_{2}.
\end{equation}
(Note ${\cal L} \ni -{\cal W}_1$.)  Our notation $\tau, L, H$ indicates
that we are thinking of these as the essentially massless tau, lepton
doublet, and down-type Higgs doublet superfields of the MSSM, with
$\lambda_{\tau}$ the usual MSSM Yukawa coupling.  Here $X_1,\; X_2,\;
Y_1,\; Y_2$ are the heavy fields which have flavor-dependent
couplings, i.e. they only couple to the $\tau$ superfield.  $h$ is a
Yukawa coupling and $M$ is a supersymmetry-preserving heavy mass.  As
discussed in the previous sections, we should add a chiral compensator,
$\Phi$, in front of mass terms and a factor
$\Phi^{\epsilon}$ in front of Yukawa couplings.  Note that all gauge
interactions have been turned off in this model.

In the second model we turn off all Yukawa couplings but add an
Abelian gauge coupling which one can think of as a new $U(1)$ flavor-dependent gauge 
interaction with gauge coupling $g'$.  The superpotential now only serves to make
the flavor fields heavy:
\begin{equation}\label{eqn:w2}
{\cal W}_2 = M \, X_{1} Y_{1} + M \, X_{2} Y_{2},
\end{equation}
and again chiral compensators must be added in front of masses.  We
keep the $\tau$ particle in the second model but drop $L$ and $H$.

The aim of this exercise is two-fold.  First of all, we have a chance 
to display how anomaly mediated calculations proceed in 
dimensional reduction.  Secondly, we will show how integrating heavy 
$X$ and
$Y$ superfields gives rise to the threshold effects that precisely
maintain the anomaly-mediation form for the scalar masses.  As
mentioned previously, this diagrammatic approach is completely
complementary to the already established approach of the spurion
calculus.  

To demonstrate the decoupling, we will calculate quantities ``above 
threshold'' and ``below threshold''.  Above threshold we are 
calculating quantities with finite external momenta well above the 
mass $M$.  In these calculations, we neglect this mass relative to 
momenta.  Below threshold, we can neglect the external momentum 
relative to the masses.  This is the energy regime where we expect to 
see the dependence on the $X$ and $Y$ vanish.

\section{$A$-Terms} \label{sec:ATerms}
In this section, we explicitly demonstrate the ultraviolet
insensitivity of the $A$-terms associated with the $\tau LH$ operator
of Equation~(\ref{eqn:w1}).  This affords us our first opportunity to
see how operators proportional to $\epsilon$ are vital to our
understanding of supersymmetry breaking in anomaly mediation.  We
calculate in bare perturbation theory and use the mass-insertion
formalism, which allows us easily to pin-point the contributions that
arise at lowest order in the gravitino mass.

Recalling Equation~(\ref{eqn:atermform}),
\begin{equation}\label{eqn:atlh}
A_{\tau LH}= - m_{3/2}\lambda_{\tau}(\gamma_{\tau} +\gamma_{L} 
+\gamma_{H}).
\end{equation}
Now, $\gamma_{\tau}$ changes as we integrate out the $X$ and $Y$
flavor superfields, and we expect to see this difference in computing
the 3-point $\tilde{\tau}\tilde{L}H$ function above and below 
threshold.  $A_{\tau L
H}$ maintains the form of Equation~(\ref{eqn:atlh}) even though its
value changes.

In the literature anomalous dimensions are typically quoted in terms
of the renormalized or running couplings $\lambda_{\tau}(\mu)$ and
$h(\mu)$ at momentum scale $\mu$.  Here we have
\begin{eqnarray}
& &\gamma_{\tau}(\mu)_{\textrm{\scriptsize Above
Threshold}}=\frac{1}{(4 \pi)^2}(2 \lambda_{\tau}
\lambda_{\tau}^{*}(\mu) + h h^*(\mu)); \label{eqn:highgamma} \\
& &\gamma_{\tau}(\mu)_{\textrm{\scriptsize Below
Threshold}}=\frac{1}{(4 \pi)^2}(2 \lambda_{\tau}
\lambda_{\tau}^{*}(\mu)),
\label{eqn:lowgamma}
\end{eqnarray}
where factors of two in front of $\lambda_{\tau}$ reflect the fact
that $L$ and $H$ are doublet fields.  To one-loop the running
couplings and bare couplings are identical, so we can freely compare
these expressions with the 3-point $\tilde{\tau}\tilde{L}H$ function computed in bare
perturbation theory.  In the two-loop scalar mass-squared computation,
however, we will need to distinguish between bare and renormalized
couplings.

In the expressions for $\gamma_{\tau}$ we see explicitly the 
ultraviolet
insensitivity: Above threshold the heavy particles contribute
$h^{*} h/(4\pi)^2$ to $\gamma_{\tau}$ or
$-m_{3/2}\lambda_{\tau}h^{*} h/(4\pi)^2$ to $A_{\tau LH}$.  Below
threshold they do not contribute at all: The soft supersymmetry-breaking
parameter $A_{\tau LH}$ is independent of the heavy-field Yukawa
parameter $h$.  Our task now is to confirm this by diagrammatic
calculation.

The relevant diagrams appear in Fig. 1.  (The mass insertion
proportional to $Mm_{3/2}$ is indicated by a cross on the $\tilde
Y\tilde X$ scalar line.)  Above threshold at scale $\mu$, {\sl
Graph~1-1} vanishes quadratically in $\frac{M^2}{\mu^2}$, so we ignore
it.  This leaves {\sl Graph~1-2} which has value
\begin{equation}\label{eq:aterm}
\mbox{\sl Graph 1-2}= i\frac{m_{3/2} h_{0}h^{*}_{0} \lambda_{\tau,0}}{(4\pi)^{2}},
\end{equation}
exactly the contribution to $A_{\tau LH}$ expected from
Equations~(\ref{eqn:atlh}) and~(\ref{eqn:highgamma}).\footnote{To keep
factors of $(-1)$ and $i$ straight, note that $i{\cal L} \ni -i
A_{\tau LH}\tilde\tau\tilde L H$ and $i{\cal L} \ni$ [{\sl Graph 1-2}]
$\tilde \tau\tilde L H$.}  As anticipated in
section~\ref{sec:regularize}, the graph with the evanescent $\epsilon$
operator produces the anomaly mediated contribution to the $A$-term.  A
graph analogous to {\sl Graph 1-2} with $L$ and $H$ fields running in
the loop contributes the $\lambda_{\tau}^{*} \lambda_{\tau}^{2}$ piece to the $A$-term
coupling.

When $\mu\ll M$, we find an additional contribution from integrating
out the $X$ and $Y$ fields, which is {\sl Graph 1-1}:
\begin{equation}
\mbox{\sl Graph 1-1}= -i\frac{m_{3/2} h_{0} h^{*}_{0} \lambda_{\tau,0}}{(4\pi)^{2}}.
\end{equation}
As promised, this is equal and opposite to the contribution from the
$\epsilon$ operator.  Together, {\sl Graph~1-1} + {\sl Graph~1-2} = 0,
so that at scales $\mu\ll M$ below threshold, the flavor-dependent 
interactions of the heavy particles do
not contribute to the $A$-term coupling.  This bears out
Equation~(\ref{eqn:lowgamma}).

It is instructive to see the dependence on the momentum scale 
$\mu^{2} = - k_{\tau}^{2}$.  The sum of {\sl Graph~1-1} and {\sl 
Graph~1-2} is 
\begin{equation}
    i \frac{m_{3/2} h_0^*h_0 \lambda_{\tau,0}}{(4\pi)^{2}}
    \left\{ 1 - \frac{4 M^{2}}{\mu \sqrt{\mu^{2}+4M^{2}}}
    \textrm{arctanh} \frac{\mu}{\sqrt{\mu^{2}+4M^{2}}} \right\},
\end{equation}
which interpolates the result above threshold (\ref{eq:aterm}) and
that below threshold (zero) as expected.

{\sl Graph~1-1} is finite by itself, so it is tempting to compute the
threshold correction without using any regulator at all.  And you do
learn something when you do this: When you compute at scales $\mu\ll
M$, you find the \emph{negative} of the expected \emph{above
threshold} ($\mu\gg M$) anomaly mediated contribution to the $A$-term.
How do we interpret this result?  This calculation computes the
correct threshold correction, but to see the ultraviolet
insensitivity, we shouldn't ignore the piece it is correcting.  A
theory is only defined after specifying a regulator, be it
Pauli-Villars, dimensional reduction, or what you will.  Thus {\sl
Graph~1-2} or its Pauli-Villars analogue always exists, regardless of
how you treat the finite {\sl Graph~1-1}.  We must regulate, and when
we include contributions from the regulator-induced operators, we find
an $A$-term which follows the trajectory defined by
Equation~(\ref{eqn:atlh}).  The regulator diagram gives the 
contribution above threshold, and {\sl Graph 1-1} gives the threshold 
correction.

\begin{figure}

\begin{picture}(200,95)(0,10)
\DashArrowArc(100,50)(40,0,180){5}
\DashArrowArc(100,50)(40,180,270){5}
\DashArrowArcn(100,50)(40,360,270){5}
\DashArrowLine(20,50)(60,50){5} 
\DashArrowLine(180,70)(140,50){5}
 \DashArrowLine(180,30)(140,50){5}
\Text(40,60)[]{$\tilde{\tau}$}
\Text(100,80)[]{$\tilde{X}$}
\Text(160,75)[]{$\tilde{L}$}
\Text(160,25)[]{$H$}
\Text(52,15)[]{$\tilde{Y}$}
\Text(142,15)[]{$\tilde{X}$}
\Text(100,10)[]{$\times$}
\Text(65,50)[l]{$hM$}
\Text(120,50)[l]{$\lambda_{\tau} h$}
\Vertex(60,50){2}
\Vertex(140,50){2}
\Text(100,0)[]{\sl Graph 1-1}
\end{picture}
\vspace{.25in}

\begin{picture}(200,95)(0,10)
\DashArrowArc(100,50)(40,0,180){5}
\DashArrowArcn(100,50)(40,360,180){5}
\DashArrowLine(20,50)(60,50){5} 
\DashArrowLine(180,70)(140,50){5}
\DashArrowLine(180,30)(140,50){5}
\Text(40,60)[]{$\tilde{\tau}$}
\Text(100,80)[]{$\tilde{X}$}
\Text(160,75)[]{$\tilde{L}$}
\Text(160,25)[]{$H$}
\Text(100,20)[]{$\tilde{X}$}
\Text(65,50)[l]{$h \epsilon m_{3/2}$}
\Text(120,50)[l]{$\lambda_{\tau} h$}
\Vertex(60,50){2}
\Vertex(140,50){2}
\Text(100,0)[]{\sl Graph 1-2}
\end{picture}
\vspace{.25in}
\caption {Diagrams that contribute to the $A_{\tau L H}$ coupling.}
\end{figure}
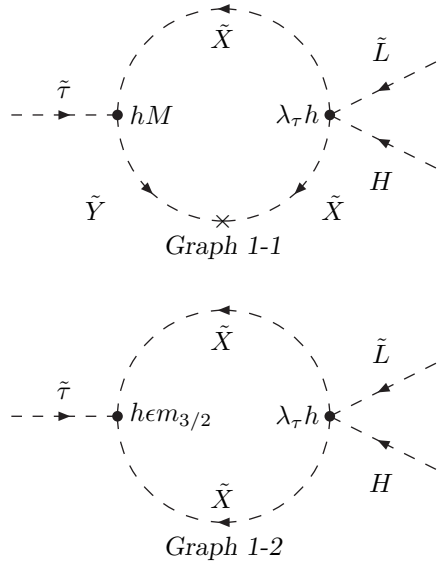

\section{Scalar Masses}\label{sec:ScalarMasses}

As a final test of our formalism, we now compute the different above and below-threshold
anomaly mediated contributions to the scalar masses.  We recover the result of ultraviolet insensitivity, providing a resolution to the puzzle of Section \ref{sec:DREDful}.  To compare
diagrammatic results with expressions for $\tilde m_{\tau}^2$, note
that the diagrams ``{\sl Graph ---}'' we compute are corrections to
$i{\cal L}$, while $-i\tilde m_{\tau}^2 \tilde{\tau}^{*}
\tilde{\tau} \in i{\cal L}$.  

\subsection{Yukawa Theory}

\subsubsection{Expectations}
To understand our diagrammatic computation, we should first work out 
what we expect. We know that the scalar masses follow the form of 
Equation~(\ref{eqn:scalarform}),
\begin{equation}
\tilde{m}^{2}_{\tau}=\frac{1}{2} m_{3/2}^2 \dot{\gamma}_{\tau}.
\end{equation}
For easy comparison with the literature, we display 
$\dot\gamma_{\tau}$ in terms of renormalized couplings.  However, we
generally work in bare perturbation theory, and at two loops the
renormalized and bare couplings differ significantly.  In particular,
when $\gamma$ and $\dot\gamma$ are written in terms of the bare
couplings, they contain additional scheme-dependent terms that vanish
in the limit that the cutoff, $\Lambda$, is taken to infinity (or
$\epsilon\to 0$).  Nevertheless, these additional terms are important
during the regularized calculation, so we re-express $\gamma$ and
$\dot{\gamma}$ in terms of bare couplings.

Working above threshold with the renormalized couplings,
\begin{eqnarray}
    & & \gamma_{\tau}(\mu)=\frac{1}{(4\pi)^2}(2 \lambda^*_{\tau}(\mu)
    \lambda_{\tau}(\mu) + h^{*}(\mu) h(\mu));\\ & &
    \gamma_{X_{1}}(\mu)=\gamma_{X_{2}}(\mu)=\frac{1}{(4\pi)^2}(h^{*}(\mu)
    h(\mu));\\ & &\dot{\gamma_{\tau}}(\mu)=\frac{1}{(4\pi)^2}(4
    \lambda_{\tau}(\mu) \dot{\lambda_{\tau}^{*}}(\mu) + 2 h(\mu)
    \dot{h^{*}}(\mu))\nonumber \\ &
    &\phantom{\dot{\gamma_{\tau}}(\mu)}= \frac{1}{(4\pi)^2}(4
    \lambda_{\tau}^*(\mu)
    \lambda_{\tau}(\mu)(\gamma_{\tau}+\gamma_L+\gamma_H) \nonumber
    \\ & & \phantom{\dot{\gamma_{\tau}}(\mu)\frac{1}{(4\pi)^2}(} + 2
    h^{*}(\mu) h(\mu)(\gamma_{\tau}+\gamma_{X_1}+\gamma_{X_2})).
\end{eqnarray}
Below threshold the terms proportional to $h^{*} h$ disappear, and in
addition, $\gamma_{\tau}$ changes as from Equation~(\ref{eqn:highgamma})
to Equation~(\ref{eqn:lowgamma}).  Expanding to pinpoint the
contributions to the scalar mass which change across the $X$ and $Y$ 
threshold, we find
\begin{eqnarray}
    \tilde{m}^{2}_{\tau} &=& \frac{m_{3/2}^2}{(4\pi)^4} (2 \lambda_{\tau}^{*} \lambda_{\tau}
    (4\lambda_{\tau}^{*} \lambda_{\tau} + h^{*} h) +
    h^{*} h (2\lambda_{\tau}^{*} \lambda_{\tau} + 3 h^{*} h)) 
    \nonumber \\
    && \qquad {(\textrm{Above
    Threshold})};\label{eqn:scalarmasshigh} \\
    \tilde{m}^{2}_{\tau} &=& \frac{m_{3/2}^2}{(4 \pi)^4}(8
    \lambda_{\tau}^{*} \lambda_{\tau} )\nonumber\\
    && \qquad{(\mbox{Below Threshold})}.\label{eqn:scalarmasslow}
\end{eqnarray}

Again, these expressions are written in terms of running couplings
$\lambda_{\tau}(\mu)$, $h(\mu)$, and in keeping with our previously
stated protocol, we now rewrite them in terms of bare couplings.  By
straight-forward computation with DRED regularization, we can compute $\log {Z}$, from which it is straight-forward to extract $\tilde{m_{\tau}}$\footnote{As 
an
alternative to direct computation, we can find $\dot\gamma_{\tau}$, and
hence $\tilde m_{\tau}^2$, through 
renormalization
group arguments.  This method is explicitly implemented for the gauge theory in an Appendix.}.  We find:
\begin{eqnarray}  \label{eqn:scalarmasshighbare}
    \lefteqn{
    \tilde{m}^{2}_{\tau} = \frac{m_{3/2}^2}{(4\pi)^2}
    \left\{-2\frac{\epsilon \lambda_{\tau,0}^{*} \lambda_{\tau,0}}{(\mu^2)^{\epsilon}} - \frac{\epsilon h^{*}_{0} h_{0}}{(\mu^2)^{\epsilon}} + \right. 
    }\nonumber \\ 
    && \left.
    \frac{1}{(4\pi)^2} \left(16
    \frac{(\lambda_{\tau,0}^{*} \lambda_{\tau,0})^2}{(\mu^2)^{2\epsilon}} +6 \frac{
    (h^{*}_{0} h_{0})^{2}}{(\mu^2)^{2\epsilon}} + 8\frac{\lambda_{\tau,0}^{*} \lambda_{\tau,0} h^{*}_{0} h_{0}}{(\mu^2)^{2\epsilon}}\right) \right\} 
    \nonumber \\ 
    && 
    {(\mbox{Above Threshold, Bare Couplings})}.
\end{eqnarray}

Below threshold, terms in $Z_{\tau}$
proportional to $h^{*}_{0} h_{0}/(\mu^2)^{\epsilon}$ are modified to
$h^{*}_{0} h_{0}/(M^2)^{\epsilon}$, because $X$ fluctuations are cut off at scales
$\mu\ll M$.  This means that most of the $h$ dependence drops out of
$\dot\gamma_{\tau}$, as in Equation~(\ref{eqn:scalarmasslow}).  Here
however, a $\lambda_{\tau,0}^{*} \lambda_{\tau,0}  h^{*}_{0} h_{0}$ term remains:
\begin{eqnarray}\label{eqn:scalarmasslowbare}
    \tilde{m}^{2}_{\tau} &=& \frac{m_{3/2}^2}{(4\pi)^2} 
    \left\{-2\frac{\epsilon
    \lambda_{\tau,0}^{*} \lambda_{\tau,0}}{(\mu^2)^{\epsilon}} \right.\nonumber \\
    & &\left.
    + \frac{1}{(4\pi)^2} \left(16 \frac{(\lambda_{\tau,0}^{*} \lambda_{\tau,0})^{2}}{(\mu^2)^{2\epsilon}} + 2 \frac{\lambda_{\tau,0}^{*} \lambda_{\tau,0} h^{*}_{0} h_{0}} {(\mu^2)^{\epsilon}(M^2)^{\epsilon}} \right) \right\} \\
    &&
    {(\mbox{Below Threshold, Bare Couplings})}.\nonumber
\end{eqnarray}
The residual $h_{0}$ dependence below threshold just reflects our use of
bare couplings.  Of course the heavy particles decouple from
the physics at scales $\mu\ll M$, and we see this when we use
renormalized couplings as in Equation~(\ref{eqn:scalarmasslow}).  As an aside, we mention that there is a factor of two difference 
between terms that go like the fourth power of the coupling constant 
when we compare Equations~(\ref{eqn:scalarmasslowbare}) and (\ref{eqn:scalarmasslow}).  The 
reason is that in Equation~(\ref{eqn:scalarmasslowbare}), part of the $(\lambda_{\tau,0}^{*} \lambda_{\tau,0})^{2}$ term combines with the ${\cal O}(\epsilon)$ piece 
to give a piece that vanishes in the four-dimensional limit.

\subsubsection{One-Loop Contributions}

We now turn to the calculation of the diagrams.  As mentioned
previously, for simplicity we compute below-threshold contributions 
to $\tilde m_{\tau}^2$ at zero
external momentum.  Above threshold we neglect the $X$ and $Y$ mass 
$M$ relative to a finite external momentum.  This procedure, together 
with the mass
insertion formalism, means that in any given diagram there is only one
fixed mass/momentum scale, a tremendous advantage computationally.
Further, when $M\to 0$, there are fewer vertices and consequently
many fewer diagrams.

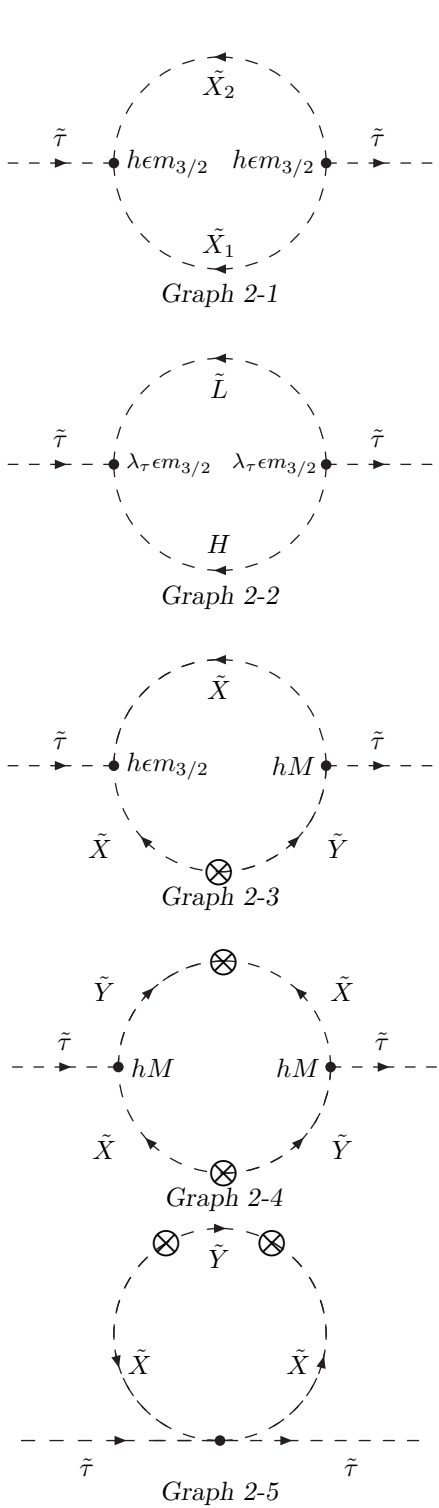
\begin{figure}\label{fig:epsoneloop}
\begin{center}

\begin{picture}(200,95)(0,10)
\DashArrowArc(100,50)(40,0,180){5}
\DashArrowArcn(100,50)(40,0,180){5}
\DashArrowLine(20,50)(60,50){5} 
\DashArrowLine(140,50)(180,50){5} 
\Text(40,60)[]{$\tilde{\tau}$}
\Text(100,80)[]{$\tilde{X_2}$}
\Text(160,60)[]{$\tilde{\tau}$}
\Text(100,20)[]{$\tilde{X_1}$}
\Text(65,50)[l]{$h \epsilon m_{3/2}$}
\Text(105,50)[l]{$h \epsilon m_{3/2}$}
\Vertex(60,50){2}
\Vertex(140,50){2}
\Text(100,0)[]{\sl Graph 2-1}
\end{picture}
\vspace{.25in}

\begin{picture}(200,95)(0,10)
\DashArrowArc(100,50)(40,0,180){5}
\DashArrowArcn(100,50)(40,0,180){5}
\DashArrowLine(20,50)(60,50){5} 
\DashArrowLine(140,50)(180,50){5} 
\Text(40,60)[]{$\tilde{\tau}$}
\Text(100,80)[]{$\tilde{L}$}
\Text(160,60)[]{$\tilde{\tau}$}
\Text(100,20)[]{${H}$}
\Text(65,50)[l]{{\footnotesize $\lambda_{\tau} \epsilon m_{3/2}$}}
\Text(105,50)[l]{{\footnotesize $\lambda_{\tau} \epsilon m_{3/2}$}}
\Vertex(60,50){2}
\Vertex(140,50){2}
\Text(100,0)[]{\sl Graph 2-2}
\end{picture}
\vspace{.25in}

\begin{picture}(200,95)(0,10)
\DashArrowArc(100,50)(40,0,180){5}
\DashArrowArcn(100,50)(40,0,90){5}
\DashArrowArc(100,50)(40,270,360){5}
\DashArrowLine(20,50)(60,50){5} 
\DashArrowLine(140,50)(180,50){5} 
\Text(40,60)[]{$\tilde{\tau}$}
\Text(100,80)[]{$\tilde{X}$}
\Text(160,60)[]{$\tilde{\tau}$}
\Text(55,20)[]{$\tilde{X}$}
\Text(145,20)[]{$\tilde{Y}$}
\Text(100,10)[]{{\Large{\bf$\bigotimes$}}}
\Text(65,50)[l]{$h\epsilon m_{3/2}$}
\Text(120,50)[l]{$h M$}
\Vertex(60,50){2}
\Vertex(140,50){2}
\Text(100,0)[]{\sl Graph 2-3}
\end{picture}
\vspace{.25in}  

\begin{picture}(200,95)(0,10)
\DashArrowArc(100,50)(40,0,90){5}
\DashArrowArcn(100,50)(40,180,90){5}
\DashArrowArcn(100,50)(40,0,90){5}
\DashArrowArc(100,50)(40,270,360){5}
\DashArrowLine(20,50)(60,50){5} 
\DashArrowLine(140,50)(180,50){5} 
\Text(40,60)[]{$\tilde{\tau}$}
\Text(55,80)[]{$\tilde{Y}$}
\Text(145,80)[]{$\tilde{X}$}
\Text(160,60)[]{$\tilde{\tau}$}
\Text(55,20)[]{$\tilde{X}$}
\Text(145,20)[]{$\tilde{Y}$}
\Text(100,90)[]{{\Large{\bf$\bigotimes$}}}
\Text(100,10)[]{{\Large{\bf$\bigotimes$}}}
\Text(65,50)[l]{$h M$}
\Text(120,50)[l]{$h M$}
\Vertex(60,50){2}
\Vertex(140,50){2}
\Text(100,0)[]{\sl Graph 2-4}
\end{picture}

\vspace{.35in}
\begin{picture}(200,95)(0,0)
\DashArrowArc(100,60)(40,270,60){5}
\DashArrowArc(100,60)(40,120,270){5}
\DashArrowArcn(100,60)(40,240,300){5}
\Text(120,94.641)[]{{\Large{\bf$\bigotimes$}}}
\Text(80,94.641)[]{{\Large{\bf$\bigotimes$}}}
\DashArrowLine(25,20)(100,20){5} 
\DashArrowLine(75,20)(175,20){5} 
\Text(50,10)[]{$\tilde{\tau}$}
\Text(150,10)[]{$\tilde{\tau}$}
\Vertex(100,20){2}
\Text(70,50)[]{$\tilde{X}$}
\Text(130,50)[]{$\tilde{X}$}
\Text(100,90)[]{$\tilde{Y}$}
\Text(100,0)[]{\sl Graph 2-5}
\end{picture}  
\vspace{.25in}
\caption{Diagrams contributing the one-loop, ${\cal O}(\epsilon)$ 
terms to
the scalar mass-squared.}

\end{center} 
\end{figure}

As seen in Equations~(\ref{eqn:scalarmasshighbare}) and
(\ref{eqn:scalarmasslowbare}), when we write the scalar mass in terms
of bare couplings there is a one-loop ${\cal O}(\epsilon)$ piece.
These one-loop ${\cal O}(\epsilon)$ terms occur diagrammatically as
shown in Fig. 2.
Above the $X$-$Y$ mass
threshold we can take $M\to 0$, so {\sl Graphs 2-3, 2-4}, and {\sl 
2-5} all
vanish, as they contain vertices $hM$ and/or $Mm_{3/2}$.  This leaves
{\sl Graph~2-1} and {\sl Graph~2-2}.  Poles from the logarithmically
divergent loop integrals pair with the ${\cal O}(\epsilon^{2})$ 
contribution from the vertices to give ${\cal O}(\epsilon)$ results:
\begin{equation}
\mbox{\sl Graph~2-1}=i\epsilon
h^{*}_{0} h_{0}\frac{m_{3/2}^{2}}{(4\pi)^{2}(\mu^2)^{\epsilon}}
\end{equation}
\begin{equation}
\mbox{\sl Graph~2-2}=2i\epsilon
\lambda_{\tau,0}^{*} \lambda_{\tau,0}\frac{m_{3/2}^{2}}{(4\pi)^{2}(\mu^2)^{\epsilon}},
\end{equation}
matching our expectations from Equation~(\ref{eqn:scalarmasshighbare}). 
Below threshold, {\sl Graph~2-1} comes with $(\mu^{2})^{\epsilon}$ 
replaced by $(M^{2})^{\epsilon}$, while
{\sl Graph~2-3, Graph 2-4} and {\sl Graph 2-5} sum to 
give
\begin{eqnarray}
\mbox{{\sl Graph~2-3} + {\sl Graph~2-4} + {\sl Graph 2-5}}= \nonumber \\
-i\epsilon
h^{*}_{0} h_{0}\frac{m_{3/2}^{2}}{(4\pi)^{2}(M^2)^{\epsilon}},
\end{eqnarray}
canceling the $h^{*}_{0} h_{0}$ dependence of Equation~(\ref{eqn:scalarmasshighbare}) 
as required by Equation~(\ref{eqn:scalarmasslowbare}).

Other one-loop graphs potentially contributing finite terms to the
scalar mass cancel among themselves.

\subsubsection{$\lambda_{\tau,0}^* \lambda_{\tau,0} h^{*}_{0} h_{0}$ Contributions}

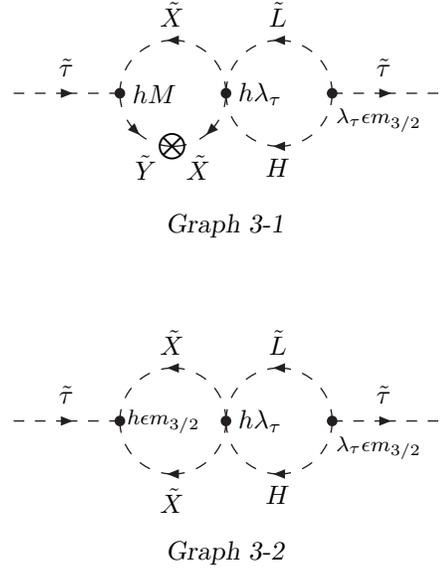
\begin{figure} \label{fig:lambdasqhsq}
\begin{center}
\begin{picture}(200,105)(0,10)
\DashArrowArc(80,60)(20,180,270){5}
\DashArrowArcn(80,60)(20,360,270){5}
\DashArrowArc(80,60)(20,360,180){5} \DashArrowArc(120,60)(20,0,180){5}
\DashArrowArcn(120,60)(20,0,180){5} \DashArrowLine(20,60)(60,60){5}
\DashArrowLine(140,60)(180,60){5} \Text(40,70)[]{$\tilde{\tau}$}
\Text(80,90)[]{$\tilde{X}$} \Text(120,90)[]{$\tilde{L}$}
\Text(160,70)[]{$\tilde{\tau}$} \Text(70,32)[]{$\tilde{Y}$}
\Text(90,32)[]{$\tilde{X}$} 
\Text(80,40)[]{{\Large{\bf$\bigotimes$}}} 
\Text(120,32)[]{${H}$}
\Vertex(100,60){2} \Vertex(60,60){2} \Vertex(140,60){2}
\Text(65,60)[l]{$hM$} \Text(105,60)[l]{$h\lambda_{\tau}$}
\Text(142,50)[l]{{\footnotesize $\lambda_{\tau}\epsilon m_{3/2}$}}
\Text(100,10)[]{\sl Graph 3-1}
\end{picture}  
\vspace{.25in}

\begin{picture}(200,105)(0,10)
\DashArrowArcn(80,60)(20,360,180){5}
\DashArrowArc(80,60)(20,360,180){5}
\DashArrowArc(120,60)(20,0,180){5}
\DashArrowArcn(120,60)(20,0,180){5}
\DashArrowLine(20,60)(60,60){5} 
\DashArrowLine(140,60)(180,60){5} 
\Text(40,70)[]{$\tilde{\tau}$}
\Text(80,90)[]{$\tilde{X}$}
\Text(120,90)[]{$\tilde{L}$}
\Text(160,70)[]{$\tilde{\tau}$}
\Text(80,30)[]{$\tilde{X}$}
\Text(120,32)[]{${H}$}
\Vertex(100,60){2}
\Vertex(60,60){2}
\Vertex(140,60){2}
\Text(63,60)[l]{{\footnotesize $h\epsilon m_{3/2}$}}
\Text(105,60)[l]{$h\lambda_{\tau}$}
\Text(142,50)[l]{{\footnotesize $\lambda_{\tau}\epsilon m_{3/2}$}}
\Text(100,10)[]{\sl Graph 3-2}
\end{picture}  
\vspace{.25in}

\caption{Diagrams that contribute $\lambda_{\tau,0}^*
\lambda_{\tau,0}^* h^{*}_{0} h_{0}$ terms to the scalar mass.}
\end{center} 
\end{figure}

There are two types of $\lambda_{\tau}^2h^2$ contributions to the
scalar mass.  The straightforward two-loop diagrams appear in Fig. 3.
Only {\sl Graph~3-2} exists above
threshold:

\begin{equation}\label{eqn:l2h2twoloop}
\mbox{\sl Graph~3-2}=-4i \lambda_{\tau,0}^{*} \lambda_{\tau,0} h^{*}_{0} h_{0}\frac{
m_{3/2}^{2}}{(4\pi)^{4}(\mu^2)^{2\epsilon}}.
\end{equation}

This is half of the $\lambda_{\tau,0}^{*} \lambda_{\tau,0} h^{*}_{0} h_{0}$ dependence needed for
$\tilde m_{\tau}^2$ in Equation~(\ref{eqn:scalarmasshighbare}).  As
expected, it is the diagram containing vertices proportional to $\epsilon$ which yields the contribution to the anomaly mediated scalar mass.

The other above-threshold contribution comes from the cross term
between the wave-function renormalization and the ${\cal O}(\epsilon)$
one-loop scalar mass derived in the previous subsection.  Since the
anomaly mediated soft scalar mass (Equation~(\ref{eqn:scalarform})
or~(\ref{eqn:scalarmasshighbare})) is the mass in a canonically
normalized Lagrangian, we need to divide the mass-renormalization part
of our two-point function by the wave-function-renormalization part,
$Z_{\tau}=1+\delta Z_{\tau}$, when computing corrections to the
mass-squared.  Cross terms between $\delta Z_{\tau}$ and two-loop mass
diagrams are higher order, but cross terms between $\delta Z_{\tau}$
and the one-loop mass diagrams contribute at ${\cal 
O}(\lambda_{\tau}^2
h^2)$.

\begin{figure}\label{fig:wavefn}
\begin{center}
\begin{picture}(200,95)(0,10)
\ArrowArc(100,50)(40,0,180)
\ArrowArcn(100,50)(40,0,180)
\DashArrowLine(20,50)(60,50){5} 
\DashArrowLine(140,50)(180,50){5} 
\Text(40,60)[]{$\tilde{\tau}$}
\Text(100,80)[]{$X_1$}
\Text(160,60)[]{$\tilde{\tau}$}
\Text(100,20)[]{$X_2$}
\Text(65,50)[l]{$h$}
\Text(125,50)[l]{$h$}
\Vertex(60,50){2}
\Vertex(140,50){2}
\Text(100,0)[]{\sl Graph 4-1}
\end{picture}
\vspace{.25in}

\begin{picture}(200,95)(0,10)
\ArrowArc(100,50)(40,0,180)
\ArrowArcn(100,50)(40,0,180)
\DashArrowLine(20,50)(60,50){5} 
\DashArrowLine(140,50)(180,50){5} 
\Text(40,60)[]{$\tilde{\tau}$}
\Text(100,80)[]{$L$}
\Text(160,60)[]{$\tilde{\tau}$}
\Text(100,20)[]{$\tilde{H}$}
\Text(65,50)[l]{$\lambda_{\tau}$}
\Text(125,50)[l]{$\lambda_{\tau}$}
\Vertex(60,50){2}
\Vertex(140,50){2}
\Text(100,0)[]{\sl Graph 4-2}
\end{picture}
\vspace{.25in}
\caption{Diagrams for the one-loop $O(1/\epsilon)$ wave-function
renormalization.}
\end{center} 
\end{figure}

Since $\delta Z_{\tau}$ will multiply the $O(\epsilon)$ one-loop
masses, we only need the $O(1/\epsilon)$ poles.  With external 
momentum $p$,
\begin{equation}
\mbox{\sl Graph~4-1}=i
h^{*}_{0} h_{0}\frac{p^{2}}{(4\pi)^{2}(-p^2)^{\epsilon}}\frac{1}{\epsilon} + 
{\cal O}(\epsilon^{0})
\end{equation}
\begin{equation}
\mbox{\sl Graph~4-2}=2i
\lambda_{\tau,0}^{*} \lambda_{\tau,0}\frac{p^{2}}{(4\pi)^{2}(-p^2)^{\epsilon}}\frac{1}{\epsilon} 
+ {\cal O}(\epsilon^{0}),
\end{equation}
which means 
\begin{equation}
\delta Z_{\tau} =
\frac{h^{*}_{0} h_{0}+2\lambda_{\tau,0}^{*} \lambda_{\tau,0}}{(4\pi)^{2}(\mu^2)^{\epsilon}}\frac{1}{\epsilon}
+ {\cal O}(\epsilon^{0}).
\end{equation}
To lowest order, dividing by $Z_{\tau}= 1+\delta Z_{\tau}$ means
multiplying by $(1-\delta Z_{\tau})$, so our sought-after contribution
is
\begin{eqnarray} \label{eqn:crossterm}
& & -\delta Z_{\tau} \times (\mbox{{\sl Graph 2-1 }+ {\sl Graph 2-2}})
\nonumber \\ &&=(-2i \lambda_{\tau,0}^{*} \lambda_{\tau,0} h^2 -2i \lambda_{\tau,0}^{*} \lambda_{\tau,0} h^{*}_{0} h_{0} -i
(h^{*}_{0} h_{0})^{2} \nonumber \\
& & -4i(\lambda_{\tau,0}^{*} \lambda_{\tau,0})^{2})\frac{
m_{3/2}^{2}}{(4\pi)^{4}(\mu^2)^{2\epsilon}}.
\end{eqnarray}
These two $2\lambda_{\tau,0}^{*} \lambda_{\tau,0}h_0^*h_0$ pieces
combines with the $4\lambda_{\tau,0}^{*} \lambda_{\tau,0}h_0^*h_0$ piece
from Equation~(\ref{eqn:l2h2twoloop}) to exhaust the
$8\lambda_{\tau,0}^{*} \lambda_{\tau,0}h_0^*h_0$ of
Equation~(\ref{eqn:scalarmasshighbare}).

Below threshold the cancellation of much of the $\lambda_{\tau,0}^{*} \lambda_{\tau,0} h_0^*h_0$
dependence proceeds as follows: we find {\sl Graph 3-1} supplies a threshold
correction 
\begin{equation}
\mbox{\sl Graph~3-1}=4i h^{*}_{0} h_{0}\lambda_{\tau,0}^{*} \lambda_{\tau,0}\frac{
m_{3/2}^{2}}{(4\pi)^{4}(\mu^2)^{\epsilon}(M^2)^{\epsilon}},
\end{equation}
which exactly cancels {\sl Graph 3-2} in 
Equation~(\ref{eqn:l2h2twoloop}) after the replacement 
$(\mu^{2})^{2\epsilon} \rightarrow (\mu^2)^{\epsilon}(M^2)^{\epsilon}$.
We already discussed in the previous subsection how {\sl Graphs 2-3,
2-4,} and {\sl 2-5} cancel {\sl Graph 2-1} below threshold.  This
leaves the cross term between $\delta Z_{\tau}$ and {\sl Graph~2-2},
one of the terms from Equation~(\ref{eqn:crossterm}).  Below threshold
the $h^{*}_{0} h_{0}(\mu^2)^{-\epsilon}$ dependence in $\delta Z_{\tau}$ becomes
$h^{*}_{0}h_{0} (M^2)^{-\epsilon}$, so that the cross term becomes
\begin{equation}
 -\delta Z_{\tau} \times (\mbox{{\sl Graph 2-2}\/})\ni
-2i\lambda_{\tau,0}^{*} \lambda_{\tau,0} h_0^*h_0\frac{
m_{3/2}^{2}}{(4\pi)^{4}(\mu^2)^{\epsilon}(M^2)^{\epsilon}},
\end{equation}
which is the residual $\lambda_{\tau,0}^{*} \lambda_{\tau,0} h^{*}_{0} h_{0}$ dependence in
Equation~(\ref{eqn:scalarmasslowbare}).  This confirms the ultraviolet
insensitivity: We have checked Equation~(\ref{eqn:scalarmasslowbare}), 
and
when we rewrite that equation in terms of renormalized couplings, we
find Equation~(\ref{eqn:scalarmasslow}).  There the ultraviolet
insensitivity is manifest.

\subsubsection{$(h^{*}_{0} h_{0})^{2}$ Contributions}

For now we continue to work exclusively with bare couplings; the
relevant $\tilde m_{\tau}^2$ for comparison is that of
Equations~(\ref{eqn:scalarmasshighbare}) 
and (\ref{eqn:scalarmasslowbare}).
The new $(h^{*}_{0} h_{0})^{2}$ diagrams appear in Fig. 5 and
Fig. 6.  The graphs shown are merely skeletons, the
true diagrams being found by adding mass insertions and the various
trilinear couplings in all possible places.

\begin{figure*}
\begin{picture}(180,100)(0,0)
\DashArrowArc(100,50)(40,0,180){5}
\DashArrowArc(100,50)(40,180,360){5}
\DashArrowLine(20,50)(60,50){5} 
\Text(80,40)[]{$\tilde{\tau}$}
\Text(40,40)[]{$\tilde{\tau}$}
\Text(160,40)[]{$\tilde{\tau}$}
\DashArrowLine(140,50)(180,50){5} 
\DashArrowLine(60,50)(130,50){5} 
\Vertex(60,50){2}
\Vertex(140,50){2}
\Text(100,0)[]{\sl Graph 5-1} 
\end{picture}
\hspace {-0.6in}
\begin{picture}(170,100)(0,0)
\DashArrowArcn(95,60)(25,180,0){5}
\DashArrowLine(20,20)(70,20){5} 
\DashArrowLine(120,20)(70,20){5}
\DashArrowLine(120,20)(170,20){5} 
\DashArrowLine(70,20)(70,60){5} 
\DashArrowLine(120,60)(120,20){5} 
\DashArrowLine(70,60)(120,60){5}
\Text(95,55)[]{$\tilde{X}$}
\Text(45,10)[]{$\tilde{\tau}$}
\Text(95,10)[]{$\tilde{X}$}
\Text(150,10)[]{$\tilde{\tau}$}
\Text(95,95)[]{$\tilde{\tau}$}
\Text(57,40)[]{$\tilde{Y}$}
\Text(127,40)[]{$\tilde{Y}$}
\Vertex(70,60){2}
\Vertex(120,60){2}
\Vertex(70,20){2}
\Vertex(120,20){2}
\Text(95,0)[]{\sl Graph 5-2}
\end{picture}  
\hspace {-0.25in}
\begin{picture}(200,95)(0,10)
\DashArrowArc(100,50)(40,0,180){5}
\DashArrowArc(80,50)(20,180,360){5}
\DashArrowLine(20,50)(60,50){5} 
\DashArrowLine(60,50)(100,50){5}
\DashArrowLine(100,50)(140,50){5} 
\DashArrowLine(140,50)(180,50){5} 
\Text(40,60)[]{$\tilde{\tau}$}
\Text(80,60)[]{$\tilde{\tau}$}
\Text(120,60)[]{$\tilde{Y}$}
\Text(160,60)[]{$\tilde{\tau}$}
\Text(100,80)[]{$\tilde{X}$}
\Text(80,25)[]{$\tilde{X}$}
\Vertex(60,50){2}
\Vertex(100,50){2}
\Vertex(140,50){2}
\Text(100,10)[]{\sl Graph 5-3}
\end{picture}  

\vspace{0.1in}
\hspace{0.3in}
\begin{picture}(150,100)(0,0)
\DashArrowArc(75,80)(20,90,270){5}
\DashArrowArc(75,80)(20,270,90){5}
\DashArrowArc(75,40)(20,90,270){5}
\DashArrowArc(75,40)(20,270,90){5}
\DashArrowLine(25,20)(75,20){5} 
\DashArrowLine(75,20)(125,20){5} 
\Text(50,10)[]{$\tilde{\tau}$}
\Text(100,10)[]{$\tilde{\tau}$}
\Vertex(75,60){2}
\Vertex(75,20){2}
\Text(30,40)[]{$\tilde{X}$}
\Text(120,40)[]{$\tilde{X}$}
\Text(30,80)[]{$\tilde{X},\tilde{\tau}$}
\Text(120,80)[]{$\tilde{X},\tilde{\tau}$}
\Text(75,0)[]{\sl Graph 5-4}
\end{picture}  
\hspace{-0.25in}
\begin{picture}(150,110)(0,0)
\DashArrowLine(35,60)(75,20){5}
\DashArrowLine(75,20)(115,60){5}
\DashArrowLine(35,60)(115,60){5}
\DashArrowArc(75,60)(40,0,180){5}
\DashArrowLine(25,20)(75,20){5} 
\DashArrowLine(75,20)(125,20){5} 
\Text(50,10)[]{$\tilde{\tau}$}
\Text(100,10)[]{$\tilde{\tau}$}
\Vertex(75,20){2}
\Vertex(35,60){2}
\Vertex(115,60){2}
\Text(42,40)[]{$\tilde{X}$}
\Text(108,40)[]{$\tilde{X}$}
\Text(75,50)[]{$\tilde{\tau}$}
\Text(75,90)[]{$\tilde{Y}$}
\Text(75,0)[]{\sl Graph 5-5}
\end{picture}  
\hspace{-0.25in}
\begin{picture}(200,100)(0,0)
\DashArrowArc(100,70)(20,90,270){5}
\DashArrowArc(100,70)(20,270,90){5}
\DashArrowArc(100,20)(30,0,90){5}
\DashArrowArc(100,20)(30,90,180){5}
\DashArrowLine(25,20)(70,20){5} 
\DashArrowLine(70,20)(130,20){5} 
\DashArrowLine(130,20)(175,20){5}
\Text(45,10)[]{$\tilde{\tau}$}
\Text(100,30)[]{$\tilde{Y}$}
\Text(155,10)[]{$\tilde{\tau}$}
\Vertex(100,50){2}
\Vertex(70,20){2}
\Vertex(130,20){2}
\Text(55,40)[]{$\tilde{X}$}
\Text(140,40)[]{$\tilde{X}$}
\Text(55,70)[]{$\tilde{X}$}
\Text(140,70)[]{$\tilde{X}$}
\Text(100,0)[]{\sl Graph 5-6}
\end{picture}  
\vspace{-.2in}
\begin{center}
\hspace{-.5in}
\begin{picture}(200,105)(0,10)
\DashArrowArc(80,60)(20,180,270){5}
\DashArrowArcn(80,60)(20,360,270){5}
\DashArrowArc(80,60)(20,360,180){5}
\DashArrowArc(120,60)(20,0,180){5}
\DashArrowArcn(120,60)(20,0,90){5}
\DashArrowArc(120,60)(20,270,360){5}
\DashArrowLine(20,60)(60,60){5} 
\DashArrowLine(140,60)(180,60){5} 
\Text(40,70)[]{$\tilde{\tau}$}
\Text(80,70)[]{$\tilde{X}$}
\Text(120,70)[]{$\tilde{X}$}
\Text(160,70)[]{$\tilde{\tau}$}
\Text(70,30)[]{$\tilde{Y}$}
\Text(90,30)[]{$\tilde{X}$}
\Text(80,40)[]{{\Large{\bf$\bigotimes$}}}
\Text(110,30)[]{$\tilde{X}$}
\Text(130,30)[]{$\tilde{Y}$}
\Text(120,40)[]{{\Large{\bf$\bigotimes$}}}
\Vertex(100,60){2}
\Vertex(60,60){2}
\Vertex(140,60){2}
\Text(100,10)[]{\sl Graph 5-7}
\end{picture}  
\end{center}

\caption{Diagrams that contribute to the $h^{4}$ threshold correction 
that exclusively include scalars.  These diagrams may be defined in 
terms of the integral $I(m,n,l)$ as defined in the text.  Where 
$\tilde{X}$ is shown, it corresponds to both ${\tilde{X}_{1}}$ and 
$\tilde{X}_{2}$, as appropriate.  Also, the three point scalar 
couplings shown here are the vertices $hM \tilde{\tau} \tilde{X} 
\tilde{Y}^{*}$. As described in the text, this vertex can be replaced 
with the $\epsilon  \tilde{\tau} \tilde{X} \tilde{X}$ vertex, 
yielding additional diagrams.}
\label{fig:h4scalars}
\end{figure*}
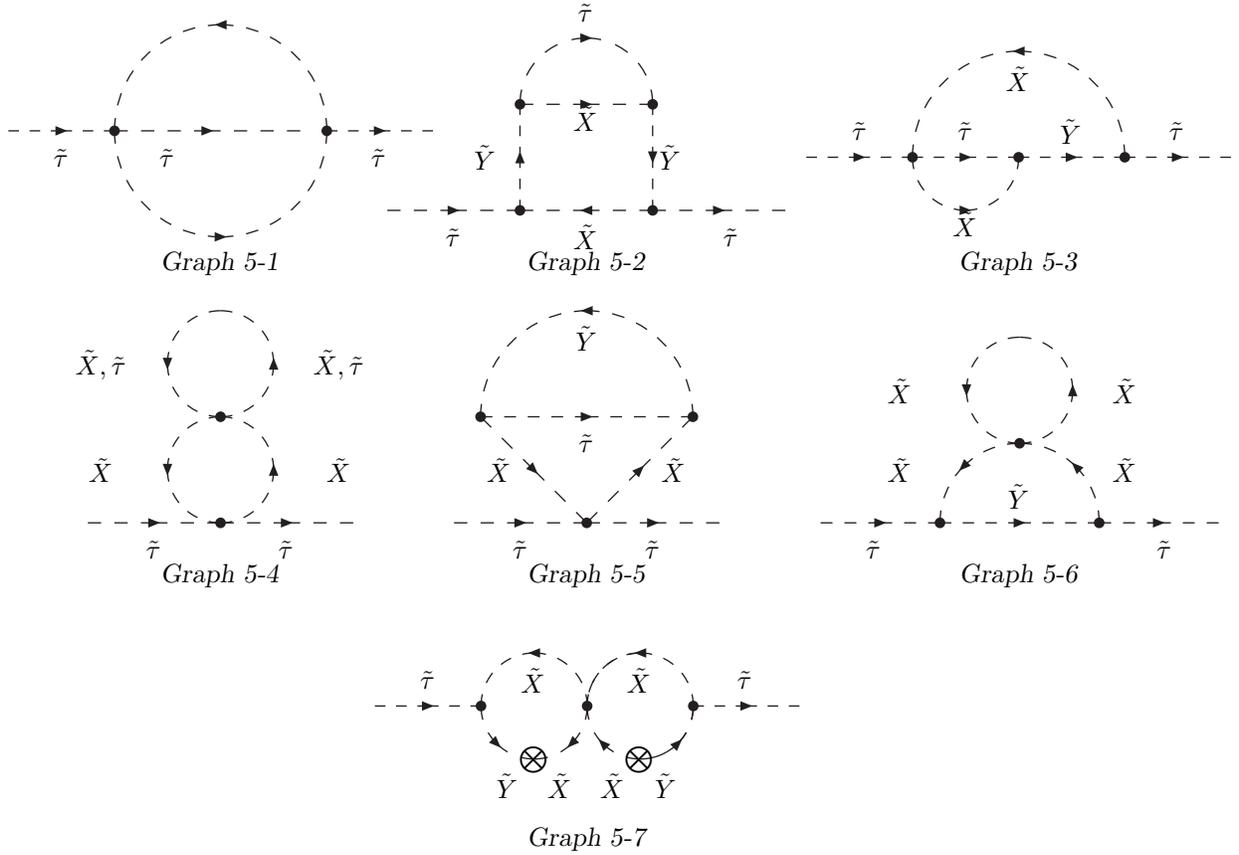

\begin{figure*}
\begin{center}
\begin{picture}(150,100)(0,0)
\DashArrowLine(35,60)(75,20){5}
\DashArrowLine(75,20)(115,60){5}
\ArrowLine(35,60)(115,60)
\ArrowArcn(75,60)(40,180,0)
\DashArrowLine(25,20)(75,20){5} 
\DashArrowLine(75,20)(125,20){5} 
\Text(50,10)[]{$\tilde{\tau}$}
\Text(100,10)[]{$\tilde{\tau}$}
\Vertex(75,20){2}
\Vertex(35,60){2}
\Vertex(115,60){2}
\Text(42,40)[]{$\tilde{X}$}
\Text(108,40)[]{$\tilde{X}$}
\Text(75,50)[]{$\tau$}
\Text(75,90)[]{X}
\Text(75,0)[]{\sl Graph 6-1}
\end{picture}  
\hspace{-.4in}
\begin{picture}(200,100)(0,0)
\ArrowArcn(95,60)(25,180,0)
\DashArrowLine(20,20)(70,20){5} 
\DashArrowLine(70,20)(120,20){5}
\DashArrowLine(120,20)(170,20){5}
\DashArrowLine(70,60)(70,20){5} 
\DashArrowLine(120,20)(120,60){5} 
\ArrowLine(70,60)(120,60)
\Text(45,10)[]{$\tilde{\tau}$}
\Text(95,30)[]{$\tilde{Y}$}
\Text(150,10)[]{$\tilde{\tau}$}
\Text(95,52)[]{X}
\Text(95,95)[]{$\tau$}
\Text(57,40)[]{$\tilde{X}$}
\Text(127,40)[]{$\tilde{X}$}
\Vertex(70,60){2}
\Vertex(120,60){2}
\Vertex(70,20){2}
\Vertex(120,20){2}
\Text(95,0)[]{\sl Graph 6-2}
\end{picture}  
\hspace{-0.4in}
\begin{picture}(200,100)(0,0)
\hspace{1in}
\DashArrowArcn(95,60)(25,180,0){5}
\DashArrowLine(20,20)(70,20){5} 
\ArrowLine(120,20)(70,20)
\DashArrowLine(120,20)(170,20){5}
\ArrowLine(70,60)(70,20) 
\ArrowLine(120,20)(120,60) 
\ArrowLine(70,60)(120,60)
\Text(45,10)[]{$\tilde{\tau}$}
\Text(95,30)[]{X}
\Text(150,10)[]{$\tilde{\tau}$}
\Text(95,52)[]{$\tau$}
\Text(95,95)[]{$\tilde{X}$}
\Text(57,40)[]{X}
\Text(127,40)[]{X}
\Vertex(70,60){2}
\Vertex(120,60){2}
\Vertex(70,20){2}
\Vertex(120,20){2}
\Text(95,0)[]{\sl Graph 6-3}
\end{picture}
\end{center}
\vspace{.1in}
\caption{Diagrams that contribute to the $h^4$ threshold correction 
that include fermions.}  
\label{fig:h4fermion}
\end{figure*}
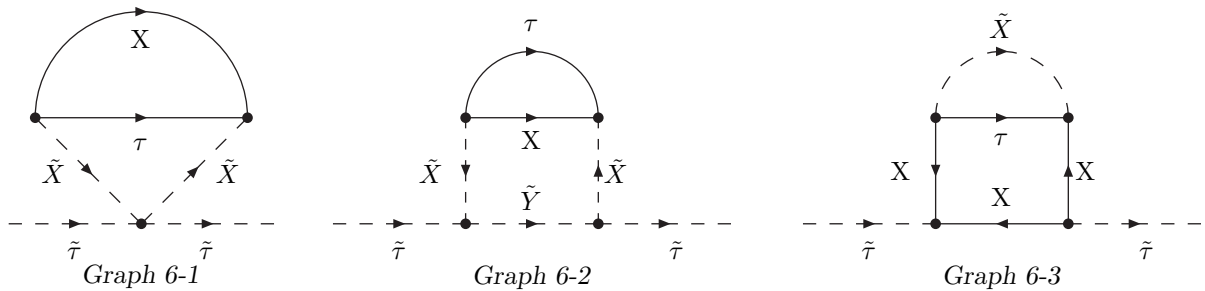

We first proceed with the calculation of the anomaly mediated
contribution to the scalar mass above threshold.  We expect our result
to agree with the $(h^{*}_{0} h_{0})^{2}$ term in
Equation~(\ref{eqn:scalarmasshighbare}). Of the graphs in the figure,
only some occur above threshold---{\sl Graphs 5-3}, {\sl 5-5}, {\sl
5-7}, and {\sl 6-2}, each with two trilinear vertices $h_{0} \epsilon
m_{3/2} \, \tilde{\tau}\tilde{X_1} \tilde{X_2}$.  The others vanish in
the $M\to 0$ limit.

The calculations are straightforward except for {\sl Graph 5-5}.  This
diagram is different from the others in that it has an infrared
divergence in the lower loop which is not regulated by the external
momentum.  However, the top loop is effectively a contribution to the
$\tilde X$ two-point function, and if we integrate that loop first, it
gives a radiatively-generated mass to the $\tilde X$ boson which
regulates the infrared divergence.  (Recall that we are working in the
limit where the tree mass $M\to 0$.  The vertices which appear in
computing the one-loop $\tilde X$ two-point function are $h_{0}\epsilon
m_{3/2} \tilde\tau\tilde X_1\tilde X_2$ and its hermitian conjugate.)  We 
will have more to say about infrared divergences in dimensional 
reduction when we discuss the gauge theory.

The values of the above-threshold diagrams appear in
Table I .  Also included is the $(h_{0}^{*} h_{0})^{2}$ contribution
derived in Equation~(\ref{eqn:crossterm}), which comes from the cross
term between the one-loop $O(\epsilon)$ scalar mass and the
wave-function renormalization.  Altogether, we find the expected
above-threshold result
\begin{equation}
\tilde m_{\tau}^2 \ni 6 (h_{0}^{*} h_{0})^{2}\frac{
m_{3/2}^{2}}{(4\pi)^{4}(\mu^2)^{2\epsilon}}.
\end{equation}
 
\begin{table} \label{table:h4above}
\begin{tabular}{lr} 
{\sl Graph 5-3} & -2 \\
{\sl Graph 5-5} & -1 \\
{\sl Graph 5-7} & -1 \\
{\sl Graph 6-2} & -1 \\
Equation~(\ref{eqn:crossterm}) & -1
\end{tabular}
\caption{Values (to ${\cal O}(\epsilon^0)$) of the diagrams suppling
above-threshold $(h^{*}_{0} h_{0})^{2}$ term in the scalar mass-squared.  We have
pulled out a common factor $\frac{1}{(4\pi)^{4}}
i (h_{0}^{*} h_{0})^{2}  m_{3/2}^{2}(\mu^2)^{-2\epsilon}$. }
\end{table}

We now turn to the calculation of the $(h_{0}^{*} h_{0})^{2}$ piece of the $\tau$-scalar mass below threshold.  Based on Equation~(\ref{eqn:scalarmasslowbare}), we expect to find zero.  Below threshold, the cross-term between the one-loop $O(\epsilon)$
mass and the wave-function renormalization disappears because the sum
of {\sl Graphs 2-3, 2-4,} and {\sl 2-5} cancels {\sl Graph 2-1}.  Then
we are left with two-loop diagrams from Fig. 5 and Fig. 6, all of
which contribute below threshold.  We split our computation into three
parts.  First, there are diagrams in which all trilinear vertices are
of the form $h_0M \tilde{\tau}\tilde{X}\tilde{Y}$, and supersymmetry-breaking comes from a pair of mass
insertions $Mm_{3/2}$ on the scalar lines.  Second, there are diagrams
with a single $\epsilon$ trilinear vertex and a single $Mm_{3/2}$
insertion.  Finally, there are the same diagrams which existed above
threshold, where two trilinear vertices are of the form $h_{0}\epsilon
m_{3/2} \tilde{\tau} \tilde{X_1} \tilde{X_2}$.  Using the integrals $I(m,n,l)$, $F(m,n,l)$, and
$G(m,n,l)$ as defined in the Appendix, we write down the values for
the Feynman diagrams in a compact form in Table II.

Expanding the integrals and summing all contributions, we find exact
cancellation, matching Equation~(\ref{eqn:scalarmasslowbare}) and
verifying ultraviolet insensitivity.  In particular, the cancellation
among the $O(\epsilon^0)$ terms looks like
\begin{equation}\label{eqn:h4cancel}
0 =i (h^{*}_{0} h_{0})^{2} \frac{m_{3/2}^{2}}{(4 \pi)^{4}(M^2)^{2\epsilon}} (-5 + 10 - 5)
\end{equation}
where the contributions are respectively from graphs with zero, one,
or two $h_0\epsilon m_{3/2} \tilde{\tau}\tilde{X_{1}}\tilde{X_{2}}$
vertices. (Table VII gives ${\cal O}(\epsilon^0)$
expansions for the integrals, but the spurion computation assures us
that the cancellation is exact, and it does indeed extend to all
orders in $\epsilon$.)

\begin{table} \label{table:h4below}
\begin{tabular}{lr} 
{\sl Graph 5-1} & $4M^2 \; I(3,1,1)$ \\
{\sl Graph 5-2} & $24M^6\; I(5,1,1)+12M^6\; I(4,2,1) +4M^6\; I(3,3,1)$ \\
{\sl Graph 5-3} & $12M^4\; I(3,2,1) + 12M^4 \; I(4,1,1)$ \\
{\sl Graph 5-4} & $4M^2 \; I(4,1,0) + 2M^2 \; I(3,2,0)$ \\ 
{\sl Graph 5-5} & $6M^4 \; I(3,2,1) + 6M^4 \; I(4,1,1)$ \\
{\sl Graph 5-6} & $2M^4 \; I(3,3,0) + 12M^4 \; I(5,1,0)$\\
{\sl Graph 5-7} & $4M^4 \; I(3,3,0)$ \\
{\sl Graph 6-1} & $-8M^2\; F(4,1,1)$ \\
{\sl Graph 6-2} & $-24M^4 \; F(5,1,1)$ \\
{\sl Graph 6-3} & $4M^6 \; G(3,3,1)$ \\
\end{tabular}
\begin{tabular}{lr} 
{\sl Graph 5-2} & $24\epsilon M^4\; I(4,1,1)+8\epsilon M^4\; I(3,2,1)$ \\
{\sl Graph 5-3} & $12\epsilon M^2\; I(3,1,1)+4\epsilon M^2\; I(2,2,1)$\\
{\sl Graph 5-5} & $4\epsilon M^2\; I(2,2,1)+4\epsilon M^2\; I(3,1,1)$ \\
{\sl Graph 5-6} & $8\epsilon M^2 \; I(4,1,0)$ \\
{\sl Graph 5-7} & $4\epsilon M^2 \; I(3,2,0)$ \\
{\sl Graph 6-2} & $-16\epsilon M^2 \; F(4,1,1)$ \\
\end{tabular}
\begin{tabular}{lr} 
{\sl Graph 5-2} & $8\epsilon^2 \; I(3,1,1)$ \\
{\sl Graph 5-3} & $4\epsilon^2 \; I(2,1,1)$ \\
{\sl Graph 5-5} & $2\epsilon^2 \; I(2,1,1)$ \\
{\sl Graph 5-6} & $2\epsilon^2 \; I(3,1,0)$ \\
{\sl Graph 5-7} & $\epsilon^2 \; I(2,2,0)$ \\
{\sl Graph 6-2} & $-4\epsilon^2 \; F(3,1,1)$ \\
\end{tabular}
\caption{Below-threshold contributions to $(h_{0}^{*} h_{0})^{2}$ terms in the scalar
mass-squared.  The three sets of values represent diagrams in which
zero, one, or two trilinear vertices are of the form $h_{0}\epsilon
m_{3/2}\tau X_1X_2$.  The integrals $I(m,n,l)$, $F(m,n,l)$, and
$G(m,n,l)$ are defined in the Appendix.  We have pulled out a common
factor $i (h_{0}^{*} h_{0})^{2} m_{3/2}^{2}$.}
\end{table}

Now it is instructive to revisit our puzzle of Section
\ref{sec:DREDful}.  When we found a vanishing threshold correction and
a resulting lack of ultraviolet insensitivity, it was because we had
not calculated all contributions to the scalar mass.  In the language
of this section, we calculated the first section of Table II, along
with a cross-term from {\sl Graph 4-1} and {\sl Graphs 2-4} and {\sl
2-5}.  We then added in a contribution from the $A$-term by hand.  This
gave an erroneous result.  We have seen that the correct procedure is
to calculate the entirety of Table II, and see that the contributions
sum to zero.

\subsubsection{Finite Computation for $(h^{*} h)^{2}$} 

\begin{figure} \label{fig:h4extraaterm}
\begin{center}
\begin{picture}(200,95)(0,10)
\DashArrowArc(100,50)(40,0,180){5}
\DashArrowArcn(100,50)(40,0,90){5}
\DashArrowArc(100,50)(40,270,360){5}
\DashArrowLine(20,50)(60,50){5} 
\DashArrowLine(140,50)(180,50){5} 
\Text(40,60)[]{$\tilde{\tau}$}
\Text(100,80)[]{$\tilde{X}$}
\Text(160,60)[]{$\tilde{\tau}$}
\Text(55,20)[]{$\tilde{X}$}
\Text(145,20)[]{$\tilde{Y}$}
\Text(100,10)[]{{\Large{\bf$\bigotimes$}}}
\Text(65,50)[l]{$h A_{\tau X X}$}
\Text(120,50)[l]{$h M$}
\Vertex(60,50){2}
\Vertex(140,50){2}
\Text(100,0)[]{\sl Graph 7-1}
\end{picture}
\vspace{.25in}  

\caption{Additional diagram for the finite $h^4$ calculation. It is
effectively a two loop diagram, as there is a one loop suppression
through the $A$-term vertex.}
\end{center} 
\end{figure}
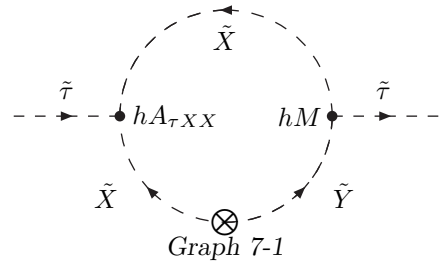

In contrast to the DRED calculation above, we present an additional calculation that does not depend on this type of regularization.  In the language of section \ref{sec:DREDful}, this calculation corresponds one where we have implicitly used a holomorphic regularization scheme.  So, we may compare this calculation to the spurion calculation done with holomorphic regularization.  This provides an additional demonstration of the ultraviolet insensitivity.  

As described in section \ref{sec:DREDful}, we must keep all integrals in four-dimensions, paying attention to the finiteness of the integrals.  
By integrating out the cut-off dependent supersymmetry-breaking operators, we recover the anomaly mediated piece of Equation~(\ref{eq:m2}).  If we choose Pauli-Villars as our holomorphic regulator, this procedure would essentially correspond to working with an effective
Lagrangian at a scale $\mu$ below the threshold of the Pauli-Villars
particles.  We have integrated out the Pauli-Villars fields, and the
anomaly mediated soft terms now appear in our effective Lagrangian.  Keeping this anomaly mediated piece in mind, we may turn to a calculation of the threshold correction.  We calculate the diagrams with $\tilde{X}$ and $\tilde{Y}$ particles in the loops, taking care to keep our integrals well defined at all times.

%


First, the $A$-terms in the effective Lagrangian give rise to the
diagram in Fig. 7.  This is effectively a
two-loop diagram because there is one-loop suppression through the
$A$-term.  It is already finite.  Recalling that $A_{\tau X_1X_2} =
-m_{3/2}h(\gamma_{\tau}+\gamma_{X_1} +\gamma_{X_2})$,
\begin{eqnarray}\label{eqn:h4extraaterm}
\mbox{\sl Graph~7-1}&=&2ih^*h\frac{
m_{3/2}^{2}}{(4\pi)^{2}}
(\gamma_{\tau}+\gamma_{X_{1}}+\gamma_{X_{2}})\\ \nonumber & =& 
2ih^*h\frac{
m_{3/2}^{2}}{(4\pi)^{4}}
(2\lambda_{\tau}^{*} \lambda_{\tau}+3h^*h).
\end{eqnarray}

The remaining relevant diagrams and their formal values appear in the
first part of Table II.  Of these only a few are
potentially divergent.  Expanding $F(l,m,n)$ and $G(l,m,n)$ in terms 
of $I(l,m,n)$, we find
\begin{equation}
\begin{array}{l}
\mbox{Sum of Diagrams}\;\;=\;\;\mbox{Finite}\;\;+  \\
\bigl( 24 M^{6} I(5,1,1) + 30 M^{4} I(4,1,1)+ 6 M^2 I(3,1,1) \bigr).
\end{array} 
\end{equation}

The last three terms are individually divergent, but their sum is
clearly not, since the left-hand side is finite.  In Appendix B, we
outline a completely finite calculation of these terms.  All told, we
find that the finite evaluation of these diagrams yields precisely
$-3i(h^{*}h)^{2} \frac{1}{(4 \pi)^{4}} m_{3/2}^{2}$.  When added with
the $6i(h^{*} h)^{2} \frac{1}{(4 \pi)^{4}} m_{3/2}^{2}$ contribution
from Equation~(\ref{eqn:h4extraaterm}), we find a total $3i(h^{*} h
)^{2}\frac{1}{(4 \pi)^{4}} m_{3/2}^{2}$.  This is the proper threshold
correction to cancel the known above-threshold contribution from the
Pauli-Villars fields as given in terms of renormalized couplings,
Equation~(\ref{eqn:scalarmasshigh}).  This again verifies ultraviolet
insensitivity.

It is worth contrasting how the cancellation happens in dimensional
reduction (Equation~(\ref{eqn:h4cancel})).  In that case the diagrams
with no $\epsilon$-dependent vertices contribute $-5(h^{*}_{0}h_{0})^{2}$, versus 
$-3(h^{*}h)^{2}$ in
the completely finite calculation.  This is a signal that we must 
include the $\epsilon$-dependent vertices to get the consistent 
results.  The dimensional reduction
cancellation happens through complicated interplay between these
diagrams and those with the new $\epsilon$ vertices.

Finally, we mention that the couplings throughout our ``finite
calculation'' are the renormalized couplings, $h(\mu)$.  This is
because we have generated the soft terms by integrating out the
Pauli--Villars fields at the cut-off scale to get
Equations~(\ref{eq:A}) and (\ref{eq:m2}).  However, these equations
are renormalization group invariant, so, we can run them down to our
threshold scale, $M$ where the equations still hold, now evaluated at
the RG-scale $M$.  This threshold correction is then done with
couplings at this scale, in other words, with the renormalized couplings.

\subsubsection{$(\lambda_{\tau,0}^{*} \lambda_{\tau,0})^{2}$ Contributions}

The calculation of $(\lambda_{\tau,0}^{*} \lambda_{\tau,0})^{2}$
contributions to $\tilde m_{\tau}^2$ is identical to the
above-threshold $(h_0^*h_0)^2$ calculation, the only difference being factors
of two from the doublets $L$ and $H$.  The table analogous to Table I
is Table III.  Summing, we find the $16(\lambda_{\tau,0}^{*}
\lambda_{\tau,0})^2$ expected in
Equation~(\ref{eqn:scalarmasshighbare}).  $(\lambda_{\tau,0}^{*}
\lambda_{\tau,0})^{2}$ contributions are not
affected by integrating out the heavy $X$ and $Y$ fields.

In summary, we have utilized our new formalism in DRED to check two
anomaly mediated calculations.  First of all, we were able to check
the usual form of the anomaly mediated contributions to $A$-terms and
scalar masses.  Secondly, we were able to explicitly verify the
ultraviolet insensitivity of anomaly mediation through a diagrammatic
calculation.

\begin{table} \label{table:lambda4above}
\begin{tabular}{lr} 
{\sl Graph 5-2} & 0 \\
{\sl Graph 5-3} & -4 \\
{\sl Graph 5-5} & -2 \\
{\sl Graph 5-7} & -4 \\
{\sl Graph 6-2} & -2 \\
Equation~(\ref{eqn:crossterm}) & -4
\end{tabular}
\caption{Values (to $O(\epsilon^0)$) of the diagrams suppling
above-threshold $(\lambda_{\tau,0}^*\lambda_{\tau,0})^2$ term in the
scalar mass-squared.  We have pulled out a common factor $i
(\lambda_{\tau,0}^*\lambda_{\tau,0})^2 \frac{1}{(4\pi)^{4}}
m_{3/2}^{2}(\mu^2)^{-2\epsilon}$.  Everywhere in {\sl Graphs 5-2}
through {\sl 6-2} $L$ and $H$ replace $X$ and $Y$.}
\end{table}

\subsection{Abelian Gauge Theory} \label{sec:gauge}

\subsubsection{Expectations}

Shifting to the U(1) gauge model described in section \ref{sec:Model},
we have particle content $\tau$, $X_1$, $Y_1$, $X_2$, $Y_2$, with
superpotential given in Equation~(\ref{eqn:w2}).  In this section we
primarily focus on additional subtleties that arise for the gauge
theory.  We show a computation of the above-threshold anomaly mediated
contributions proportional to $Y_{\tau}^{2}$ and $Y_{\tau}^{2}
Y_{X}^{2}$.  We further check that the contributions going like
$Y_{\tau}^{2} Y_{X}^{2}$ vanish below threshold, confirming
ultraviolet insensitivity.  We believe these calculations capture the
subtleties associated with the gauge theory.  Incidentally, the
calculation of the anomaly mediated contributions in this model is
quite similar to a gauge mediation calculation performed previously
\cite{gaugemed}.

Before calculating any diagrams, it is important to know what we
expect for the scalar mass.  For this, we need to know $\dot{\gamma}$.
It is useful to write the results in terms of both renormalized and
bare couplings.  In terms of renormalized couplings we have:
\begin{eqnarray}\label{eqn:gammataugauge}
\gamma_{\tau}(\mu)&=&\frac{1}{(4\pi)^2}(-2 g'^2(\mu)\, Y_{\tau}^2 );\\
\dot{\gamma_{\tau}}(\mu)&=&\frac{1}{(4\pi)^2}(-4 g'(\mu) \dot g'(\mu)
Y_{\tau}^2)\nonumber \\ &=& \frac{1}{(4\pi)^4}(-4 g'^4(\mu) Y_{\tau}^2
(Y_{\tau}^2 + Y_{X_1}^2 + Y_{Y_1}^2+ Y_{X_2}^2+ Y_{Y_2}^2))\nonumber
\\ &=& \frac{1}{(4\pi)^4}(-4 g'^4(\mu) Y_{\tau}^2 (Y_{\tau}^2 +
\sum_{\rm heavy} Y_{i}^2)).
\end{eqnarray}
(Here and below, the sum over heavy multiplets is performed for each 
chiral superfield separately.)  Then clearly, 
\begin{eqnarray}
\tilde{m}^{2}_{\tau} &=& \frac{m_{3/2}^2}{(4\pi)^4}(-2 g'^4(\mu)\,
Y_{\tau}^2 (Y_{\tau}^2 + \sum_{\rm heavy} Y_{i}^2)) \label{eqn:gaugescalarmasshigh}\\
&& {(\mbox{Above Threshold})};\nonumber\\
\tilde{m}^{2}_{\tau} &=& \frac{m_{3/2}^2}{(4 \pi)^4} (-2 
{g'^{4}(\mu)}\,
 Y_{\tau}^4) \label{eqn:gaugescalarmasslow}
\\
&& {(\mbox{Below Threshold})}.\nonumber
\end{eqnarray}
The last expression does not depend on the properties of heavy 
particles at all, manifesting the UV insensitivity.  To get the analogous expressions in terms of the bare couplings 
requires a bit more work.  This calculation in done in an Appendix.  
In terms of bare couplings we have:
\begin{eqnarray}
\label{eqn:gaugescalarmasshighbare}
\tilde{m}^{2}_{\tau} &=& \frac{m_{3/2}^2}{(4\pi)^2} \left(2 \frac{\epsilon
Y_{\tau}^{2} {g'}^{2}_{0}}{ (\mu^2)^{\epsilon}} -4\frac{{g'_{0}}^{4}
Y_{\tau}^2 (Y_{\tau}^2 + \sum_{\rm heavy} Y_{i}^2)}{(4\pi)^2
(\mu^2)^{2\epsilon}}\right) \\ && {(\mbox{Above Threshold, Bare Couplings
})}.\nonumber
\\
\label{eqn:gaugescalarmasslowbare}
\tilde{m}^{2}_{\tau} &=& \frac{m_{3/2}^2}{(4\pi)^2} \left(2 \frac{\epsilon
Y_{\tau}^{2} {g'_{0}}^{2}}{ (\mu^2)^{\epsilon}} \right. \nonumber \\
& & \left. -4\frac{{g'_{0}}^{4}
Y_{\tau}^4}{(4\pi)^2
(\mu^2)^{2\epsilon}} -2\frac{{g'_{0}}^{4} Y_{\tau}^2 \sum_{heavy}
Y_{i}^2}{(4\pi)^2 (\mu^2)^{\epsilon}(M^2)^{\epsilon}}\right) \\ 
& &{(\mbox{Below Threshold,
Bare Couplings })}.\nonumber
\end{eqnarray}

\subsubsection{Insensitivity}

In this section, we will compute the above-threshold anomaly mediated
contributions proportional to $Y_{\tau}^2$ and $Y_{\tau}^{2}
Y_{X_{i}}^{2}$, and we check that the latter vanish below threshold to
confirm ultraviolet sensitivity. The relevant skeleton diagrams are shown
in Fig. 8; we must add appropriate supersymmetry breaking vertices
to form the actual diagrams.  (There are many additional diagrams
which give $Y_{\tau}^{4}$ contributions, but we do not expect further
conceptual difficulties in their calculation.)  

\begin{figure*}
\label{fig:gfourth}
\begin{center}
\begin{picture}(100,120)(0,0)
\DashArrowArcn(50,80)(20,90,270){5}
\DashArrowArcn(50,80)(20,270,90){5}
\DashArrowLine(20,20)(50,20){5} 
\DashArrowLine(50,20)(90,20){5}
\PhotonArc(50,40)(20,90,270){3}{6.5} 
\PhotonArc(50,40)(20,270,450){3}{6.5}
\Text(25,10)[]{$\tilde{\tau}$}
\Text(75,10)[]{$\tilde{\tau}$}
\Text(50,85)[]{$\tilde{X},\tilde{Y}$}
\Vertex(50,60){2}
\Vertex(50,20){2}
\Text(50,0)[]{\sl Graph 8-1}
\end{picture}
\hspace{-.25in}
\begin{picture}(170,120)(0,0)
\DashArrowArcn(95,80)(25,0,180){5}
\DashArrowArcn(95,80)(25,180,360){5}
\DashArrowLine(20,20)(70,20){5} 
\DashArrowLine(70,20)(120,20){5}
\DashArrowLine(120,20)(170,20){5}
\Photon(70,20)(70,80){3}{6.5} 
\Photon(120,80)(120,20){3}{6.5} 
\Text(45,10)[]{$\tilde{\tau}$}
\Text(95,30)[]{$\tilde{\tau}$}
\Text(150,10)[]{$\tilde{\tau}$}
\Text(95,85)[]{$\tilde{X},\tilde{Y}$}
\Vertex(70,80){2}
\Vertex(120,80){2}
\Vertex(70,20){2}
\Vertex(120,20){2}
\Text(95,0)[]{\sl Graph 8-2}
\end{picture}
\end{center}
\hspace{1in}
\begin{picture}(180,100)(0,0)
\DashArrowArc(100,50)(40,0,180){5}
\DashArrowArc(100,50)(40,180,360){5}
\DashArrowLine(20,50)(60,50){5} 
\Text(95,40)[]{$\tilde{\tau}$}
\Text(40,40)[]{$\tilde{\tau}$}
\Text(160,40)[]{$\tilde{\tau}$}
\DashArrowLine(140,50)(180,50){5} 
\DashArrowLine(60,50)(130,50){5} 
\Vertex(60,50){2}
\Vertex(140,50){2}
\Text(100,0)[]{\sl Graph 8-3} 
\end{picture}
\hspace{-.6in}
\begin{picture}(170,120)(0,0)
\DashArrowArcn(95,80)(20,90,270){5}
\DashArrowArcn(95,80)(20,270,90){5}
\DashArrowLine(20,20)(70,20){5} 
\DashArrowLine(70,20)(120,20){5}
\DashArrowLine(120,20)(170,20){5}
\Photon(70,20)(95,60){3}{4.5} 
\Photon(95,60)(120,20){3}{4.5} 
\Text(45,10)[]{$\tilde{\tau}$}
\Text(95,30)[]{$\tilde{\tau}$}
\Text(150,10)[]{$\tilde{\tau}$}
\Text(95,85)[]{$\tilde{X},\tilde{Y}$}
\Vertex(95,60){2}
\Vertex(70,20){2}
\Vertex(120,20){2}
\Text(95,0)[]{\sl Graph 8-4}
\end{picture}
\begin{center}
\begin{picture}(150,100)(0,0)
\DashArrowArc(75,80)(20,90,270){5}
\DashArrowArc(75,80)(20,270,90){5}
\DashArrowArc(75,40)(20,90,270){5}
\DashArrowArc(75,40)(20,270,90){5}
\DashArrowLine(25,20)(75,20){5} 
\DashArrowLine(75,20)(125,20){5} 
\Text(50,10)[]{$\tilde{\tau}$}
\Text(100,10)[]{$\tilde{\tau}$}
\Vertex(75,60){2}
\Vertex(75,20){2}
\Text(30,40)[]{$\tilde{X},\tilde{Y},\tilde{\tau}$}
\Text(120,40)[]{$\tilde{X}, \tilde{Y}, \tilde{\tau}$}
\Text(30,80)[]{$\tilde{X},\tilde{Y}, \tilde{\tau}$}
\Text(120,80)[]{$\tilde{X},\tilde{Y}, \tilde{\tau}$}
\Text(75,0)[]{\sl Graph 8-5}
\end{picture}  
\hspace{-0.5in}
\begin{picture}(200,100)(0,0)
\DashArrowArcn(95,60)(25,180,0){5}
\DashArrowLine(20,20)(70,20){5} 
\ArrowLine(70,20)(120,20)
\DashArrowLine(120,20)(170,20){5}
\ArrowLine(70,20)(70,60) 
\ArrowLine(120,60)(120,20)
\ArrowLine(120,60)(70,60)
\Text(45,10)[]{$\tilde{\tau}$}
\Text(95,30)[]{$\tau$}
\Text(150,10)[]{$\tilde{\tau}$}
\Text(95,52)[]{X} 
\Text(95,95)[]{$\tilde{X}$}
\Photon(70,20)(70,60){3}{3.5}
\Photon(120,20)(120,60){3}{3.5}
\Vertex(70,60){2}
\Vertex(120,60){2}
\Vertex(70,20){2}
\Vertex(120,20){2}
\Text(95,0)[]{\sl Graph 8-6}
\end{picture}
\hspace{-0.8in}
\begin{picture}(170,120)(20,0) 
\DashArrowArc(95,75)(28,210,330){5}
\DashArrowLine(40,20)(95,20){5} 
\DashArrowLine(95,20)(150,20){5}
\DashArrowArc(95,44)(28,30,150){5}
\Photon(95,20)(70,60){3}{4.25} 
\Photon(95,20)(120,60){-3}{4.25} 
\Text(45,10)[]{$\tilde{\tau}$}
\Text(150,10)[]{$\tilde{\tau}$}
\Text(95,85)[]{$\tilde{X},\tilde{Y}$}
\Vertex(70,60){2}
\Vertex(120,60){2}
\Vertex(95,20){2}
\Text(95,0)[]{\sl Graph 8-7}
\end{picture}
\end{center}
\vspace{0.1in}
\caption{Diagrams that contribute to the cancellation in 
$g\prime^{4}$.  These diagrams all contain the heavy fields $\tilde{X}$ and $\tilde{Y}$.} 
\end{figure*}
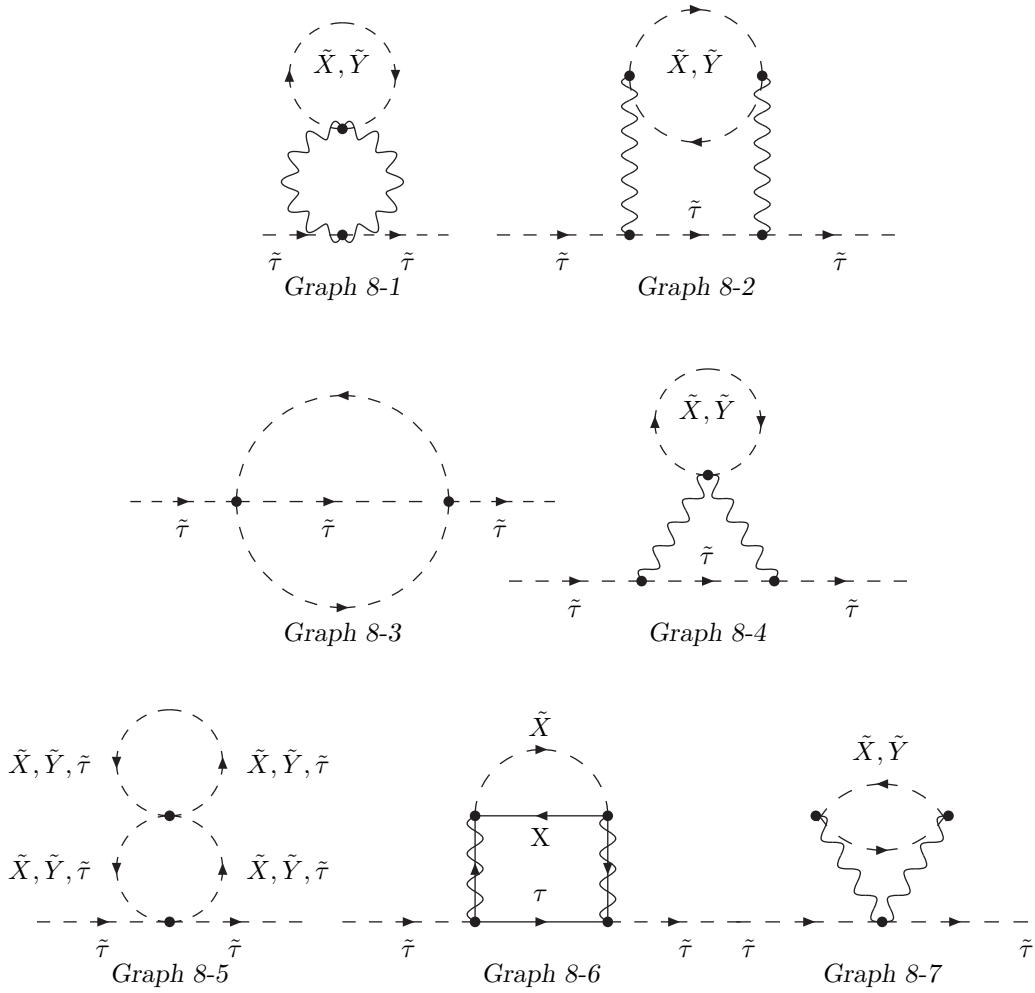

Let us consider the contribution to the scalar masses above threshold.
In this energy regime the SUSY-breaking $Mm_{3/2}$ mass insertion is suppressed ($M\to 0$), so
there are only two sources of supersymmetry-breaking.  First there is a
tree-level gaugino mass, $m_{\lambda}=-\epsilon m_{3/2}$.  Then
depending on the choice of the bare Lagrangian,
Equation~(\ref{eqn:nonlocal}) or Equation~(\ref{eqn:GMZ}), the remaining
supersymmetry-breaking is given by the non-local gaugino operator in 
Equation~(\ref{eqn:nonlocalL}) or by the $\epsilon$-scalar mass,  
$m_{\epsilon}=\epsilon m_{3/2}^{2}$ that results from using the GMZ operator.

The diagrams in Figure 9 yield the one-loop ${\cal O}(\epsilon)$ piece
in Equations~(\ref{eqn:gaugescalarmasshighbare}) and
(\ref{eqn:gaugescalarmasslowbare}).  Depending on our form for the bare
Lagrangian, either {\sl Graph 9-1} or {\sl Graph 9-2} contributes.
The values of the diagrams appear in Table IV.

\begin{figure}
\begin{center}
\vspace{.35in}
\begin{picture}(200,95)(0,0)
\DashArrowArc(100,60)(40,0,180){2}
\DashArrowArc(100,60)(40,180,0){2}
\DashArrowLine(25,20)(100,20){5} 
\DashArrowLine(75,20)(175,20){5} 
\Text(50,10)[]{$\tilde{\tau}$}
\Text(150,10)[]{$\tilde{\tau}$}
\Vertex(100,20){2}
\Text(100,90)[]{$\epsilon$-Scalar}
\Text(100,0)[]{\sl Graph 9-1}
\end{picture} 

\begin{picture}(200,95)(0,10)
\ArrowArc(100,50)(40,180,360)
\ArrowArcn(100,50)(40,180,270)
\ArrowArcn(100,50)(40,270,360)
\PhotonArc(100,50)(40,0,90){4}{4.5}
\PhotonArc(100,50)(40,90,180){4}{4.5}
\DashArrowLine(20,50)(60,50){5} 
\DashArrowLine(140,50)(180,50){5} 
\Text(40,60)[]{$\tilde{\tau}$}
\Vertex(100,90){2}
\Text(100,80)[]{$\lambda$}
\Text(160,60)[]{$\tilde{\tau}$}
\Text(100,20)[]{$\tau$}
\Vertex(60,50){2}
\Vertex(140,50){2}
\Text(100,0)[]{\sl Graph 9-2}
\end{picture}
\vspace{.25in}

\begin{picture}(200,95)(0,10)
\ArrowArc(100,50)(40,180,360)
\ArrowArcn(100,50)(40,180,120)
\ArrowArc(100,50)(40,60,120)
\ArrowArcn(100,50)(40,60,0)
\PhotonArc(100,50)(40,0,60){4}{2.5}
\PhotonArc(100,50)(40,60,120){4}{2.5}
\PhotonArc(100,50)(40,120,180){4}{2.5}
\DashArrowLine(20,50)(60,50){5} 
\DashArrowLine(140,50)(180,50){5} 
\Text(40,60)[]{$\tilde{\tau}$}
\Text(120,84.6)[]{{\Large{\bf$\bigotimes$}}}
\Text(80,84.6)[]{{\Large{\bf $\bigotimes$}}}
\Text(100,80)[]{ $\lambda$}
\Text(160,60)[]{$\tilde{\tau}$}
\Text(100,20)[]{$\tau$}
\Vertex(60,50){2}
\Vertex(140,50){2}
\Text(100,0)[]{\sl Graph 9-3}
\end{picture}
\end{center}
\vspace{.25in}
\caption{Diagrams that contribute to the ${\cal O}(\epsilon)$ scalar
mass.  {\sl Graph 9-1} exists when we utilize the GMZ operator as our
bare Lagrangian. {\sl Graph 9-2} exists when we utilize the non-local
operator: the vertex depicted in this graph is the non-local vertex of
Equation~(\ref{eqn:barenonlocal}).  {\sl Graph 9-3} exists in either 
case,
and gets its supersymmetry-breaking from the gaugino mass.}
\end{figure}
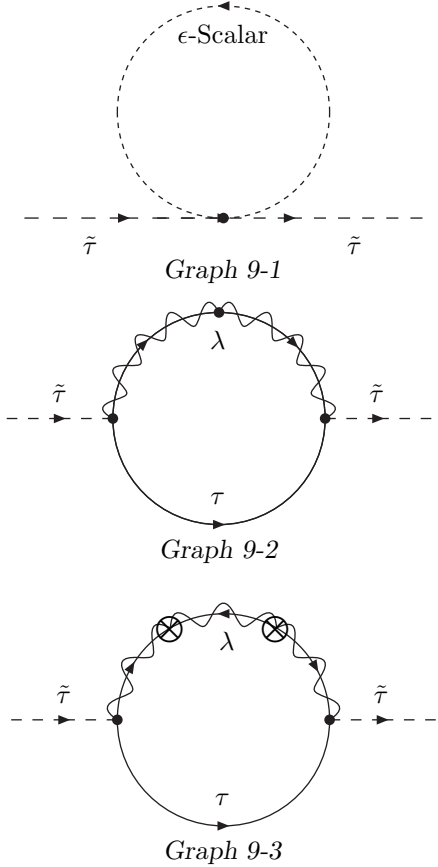

\begin{table}
\begin{tabular}{lr} 
{\sl Graph 9-1} & 2 $\epsilon$ \\
{\sl Graph 9-2} & 2 $\epsilon$\\
{\sl Graph 9-3} & -4 $\epsilon$ 
\end{tabular}
\caption{The one-loop ${\cal O}(\epsilon)$ contributions to the scalar
mass.  We have factored out the quantity $i
{g^{\prime}_{0}}^{2}\frac{1}{(4 \pi)^{2}} Y_{\tau}^{2}
m_{3/2}^{2}(\mu^2)^{-\epsilon}$. Only one of {\sl Graph 9-1} or {\sl
9-2} contributes, depending on the form of the bare Lagrangian.  The
sum of {\sl Graph 9-1} and {\sl Graph 9-3} yields a total contribution
which agrees with Equation~(\ref{eqn:gaugescalarmasshighbare}).}
\end{table} 

The two-loop terms come from one or two diagrams.  If we choose to
work with the non-local operator, only gauginos have supersymmetry-breaking,
and so only the single topology {\sl Graph 8-6} contributes.  If instead we work with a supersymmetry-breaking mass for the epsilon scalar, {\sl
Graph 10-1} adds to a reduced contribution from {\sl Graph 8-6}.
Table V collects these contributions to the scalar mass, which total
\begin{equation}
\tilde m_{\tau}^2 \ni -4 {g_{0}^{\prime}}^4 Y_{\tau}^{2}\sum_{\rm heavy}Y_i^2 
\frac{m_{3/2}^{2}}{(4
\pi)^{4} (\mu^2)^{2\epsilon}}.
\end{equation}

\begin{table}
\label{table:g4above}
\begin{tabular}{lr} 
{\sl Graph 8-6} &  6  \\
{\sl Graph 8-6} Non-Local & -2  \\
{\sl Graph 10-1} &  -2  \\
\end{tabular}
\caption{Values to $O(\epsilon^0)$ of the graphs contributing to the
${g^{\prime}_0}^{4}$ correction to the scalar mass above threshold.
We have omitted a common factor of $i {g^{\prime}_{0}}^{4}
\frac{1}{(4\pi)^{4}}m_{3/2}^2 Y_{\tau}^2\sum_{\rm heavy}Y_i^2
(\mu^2)^{-2\epsilon}$.}
\end{table}

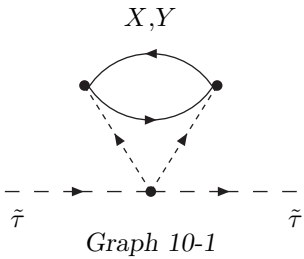
\begin{figure}
\begin{center}
\begin{picture}(170,120)(20,0) 
\ArrowArc(95,75)(28,210,330)
\DashArrowLine(40,20)(95,20){5} 
\DashArrowLine(95,20)(150,20){5}
\ArrowArc(95,44)(28,30,150)
\DashArrowLine(95,20)(70,60){2} 
\DashArrowLine(95,20)(120,60){2} 
\Text(45,10)[]{$\tilde{\tau}$}
\Text(150,10)[]{$\tilde{\tau}$}
\Text(95,85)[]{$X$,$Y$}
\Vertex(70,60){2}
\Vertex(120,60){2}
\Vertex(95,20){2}
\Text(95,0)[]{\sl Graph 10-1}
\end{picture}
\end{center}
\vspace{.25in}
\caption{$\epsilon$-scalar diagram that contributes to the 
${g_0^{\prime}}^{4}$ contribution to the scalar mass.  The 
supersymmetry-breaking arises from the $\epsilon$ scalar mass.}
\label{fig:escalar}
\end{figure}
This completes the calculation of the mass above the threshold.  We now demonstrate decoupling.  All Graphs in Fig. 8 are relevant,
because below threshold we keep a finite $X$-$Y$ mass $M$, and
supersymmetry-breaking enters through a $\tilde{X}\tilde{Y}$ mass insertion.
First we consider only the diagrams with this sort of supersymmetry-breaking.
There appear to be seven such diagrams, but several of these in fact
do not contribute.

{\sl Graph 8-5} vanishes because it is proportional to the sum of the
flavor charges of the heavy fields.  This sum vanishes by the gauge
invariance of the Lagrangian.  The sum of {\sl Graph 8-2} and {\sl
Graph 8-4} also vanishes by gauge invariance:  If we add {\sl Graph
8-4} and {\sl Graph 8-2}, we get a graph that contains the vacuum
polarization operator for scalar QED, with a form fixed by gauge
invariance to be
\begin{equation}
\Pi^{\mu\nu}=( p^{\mu}
p^{\nu}- p^{2} g^{\mu \nu}) \Pi (p^{2}).
\end{equation}
Upon contraction with the momentum-dependent $(\tilde{\tau}
\partial_{\mu}\tilde{\tau}^{*} A^{\mu}+ \textrm{h.c})$ vertex, the sum
of {\sl Graphs 8-2} and {\sl 8-4} yields zero.  Thus only four
graphs containing supersymmetry-breaking due to the $M m_{3/2}$ mass insertion contribute to the
threshold correction: {\sl Graphs 8-3}, {\sl 8-6}, and the
combination of {\sl Graph 8-1} and {\sl Graph 8-7} which again
contains the vacuum polarization operator.

There is a remaining worry concerning infrared divergences.  We can safely
express {\sl Graph 8-3} in terms of the standard integrals in the
appendix, but we must be more careful with the other graphs.  If
blithely written in terms of $I(m,n,l)$, the diagrams contain
infrared-divergent integrals which are not automatically regulated by
DRED.  While in DREG one can analytically continue to $4+2 \epsilon$
dimensions to regulate the IR, DRED by definition \emph{compactifies}
$4$ dimensions down to $4-2\epsilon$ dimensions.  IR-divergent 
integrals are thus not well
defined by DRED, and one can find mutually inconsistent ways to
evaluate such integrals.  In particular, the sometimes-seen
prescription
\begin{equation}
\int \frac{d^4p}{p^4} = 0 \qquad\qquad (\textrm{inconsistent})
\end{equation}
leads to inconsistent results.  Nonetheless, in supersymmetry we must use DRED,
and not DREG, because an extension above 4 dimensions changes the
spinor algebra and causes a mismatch in the fermionic and boson
degrees of freedom.  In short, a safe and consistent procedure is to use DRED
to regulate the UV and add finite masses to regulate the IR when
necessary.

Fortunately in our case, we can largely avoid the infrared
divergences.  There are only two cases where we find IR divergences to
be an issue.  The first is {\sl Graph 8-1} with the $\epsilon$-scalar
replaces of the vector boson.  The second is in {\sl Graph 8-6} when
there is supersymmetry breaking from the non-local contribution to the
gaugino propagator.  In these cases, we keep a finite mass.  Among the
other graphs, once {\sl Graph 8-1} and {\sl Graph 8-7} are combined,
the sum is manifestly infrared finite.  {\sl Graph 8-6}
(without the non-local term in the gaugino propagator)
and {\sl Graph 8-3} are each infrared finite on their own.  We evaluate
these two graphs directly, and their result is shown in Table VI.

We evaluate {\sl Graph 8-1} and {\sl Graph 8-7} by summing their top
loops into the vacuum polarization operator and then contracting this
subgraph with the seagull vertex.  This avoids all ambiguities due to
infrared divergences.  To compute the explicit form of the vacuum
polarization operator, we found it easier to work in the mass
eigenbasis, where the scalars have masses $M^{2}\pm M m_{3/2}$, and
then to expand to $O(m_{3/2}^2)$.  For a scalar particle of mass M and
charge $Y_{X}$ (in the mass eigenbasis),
\begin{eqnarray} \label{eqn:vacuumpolar}
\Pi(p^2)=( p^{\mu} p^{\nu}- p^{2} g^{\mu \nu}) \frac{-Y_{X}^{2} 
g_{0}^{\prime 2} i \Gamma(\epsilon)} {(4 \pi)^{2-\epsilon}} \, \times 
\nonumber \\
\int\limits_{0}^{1} { d x \, \frac{(1-2x)^{2}}{(M^{2} -p^{2} x 
(1-x))^{\epsilon}}}
\end{eqnarray}
Summing over the two eigenmasses and expanding in $m_{3/2}$, we find
the vacuum polarization in the mass insertion formalism:
\begin{eqnarray} \label{eqn:vacuumpolarMI} 
\Pi(p^2)=( p^{\mu} p^{\nu}- p^{2} g^{\mu \nu}) \times \nonumber \\
 \Bigl(i m_{3/2}^{2} M^{2}\Bigr) \frac{Y_{{X}_i}^{2} g_{0}^{\prime 2}
 \Gamma(2+ \epsilon)}{(4 \pi)^{2-\epsilon}} \times \nonumber \\
 \int\limits_{0}^{1} {dx \, \frac{(1-2 x)^{2}}{(M^{2} - p^{2} (1-x)
 x)^{2+ \epsilon}}}
\end{eqnarray} 
We contract this result with the seagull vertex to obtain a final
value for {\sl Graphs 8-1} and {\sl 8-7}.  The result appears in Table VI.

There are further contributions in which supersymmetry breaking does not come
from the B-type mass.  The tree-level gaugino mass equal to $-\epsilon
m_{3/2}$ enters a diagram identical to {\sl Graph 8-6}, but with one or both of  the
$\tilde{X} \tilde{Y}$ mass insertions replaced by gaugino mass insertions. Finally, there are the diagrams involving
either the non-local correction to the gaugino propagator
(Equation~(\ref{eqn:barenonlocal})), or a massive $\epsilon$ scalar
(Equation~(\ref{eqn:GMZ})).  For these diagrams, as mentioned above, we are careful to keep a finite mass to deal with the infrared.  

Summing the first six and either of the last two 
contributions from Table VI, we find below-threshold scalar mass
dependence
\begin{equation}
\tilde m_{\tau}^2 \ni -2 {g_{0}^{\prime}}^4 Y_{\tau}^{2}\sum_{\rm heavy}Y_i^2
\frac{m_{3/2}^{2}}{(4 \pi)^{4} (\mu^2)^{\epsilon}(M^2)^{\epsilon}}.
\end{equation}
This establishes Equation~(\ref{eqn:gaugescalarmasslowbare}), which is
in turn equivalent to Equation~(\ref{eqn:gaugescalarmasslow}), making
ultraviolet insensitivity explicit.

\begin{table}
\label{table:g4below}
\begin{tabular}{lr} 
{\sl Graph 8-7} + {\sl 8-1} &  $(\frac{3}{\epsilon} - 5 - 6 
\gamma)(M^2)^{-2\epsilon}  $  \\
Graph 8-1 ($\epsilon$-Scalar) &   $2 
(\mu^2)^{-\epsilon}(M^2)^{-\epsilon}$ \\ 
{\sl Graph 8-3} & $2 I(3,1,1) -  I(2,2,1) $\\
& $=(\frac{1}{\epsilon}-1 - 2\gamma) (M^2)^{-2 \epsilon}$ \\ 
{\sl Graph 8-6} & $(-\frac{4}{\epsilon} +2 + 8\gamma) (M^2)^{-2 \epsilon} $
\\ \hline \hline
{\sl Graph 8-6} (One $m_{\lambda}$) &
$-8(\mu^2)^{-\epsilon}(M^2)^{-\epsilon} + 8(M^2)^{-2 \epsilon} $ \\ 
{\sl Graph 8-6} (Two $m_{\lambda}$) &  
$12(\mu^2)^{-\epsilon}(M^2)^{-\epsilon} - 6(M^2)^{-2 \epsilon} $ \\ 
\hline \hline 
{\sl Graph 8-6} (Non-Local) &  
$-4(\mu^2)^{-\epsilon}(M^2)^{-\epsilon}+2(M^2)^{-2 \epsilon} $ \\ 
{\sl Graph 10-1} (GMZ) & $-4(\mu^2)^{-\epsilon}(M^2)^{-\epsilon}+2 (M^2)^{-2 
\epsilon}$
\end{tabular}
\caption{Values to $O(\epsilon^0)$ of the graphs contributing to the
${g_0^{\prime}}^{4}$ correction to the scalar mass below threshold.  We have
omitted a common factor of $i {g_0^{\prime}}^{4}\frac{1}
{(4\pi)^{4}}
Y_{\tau}^2\sum_{\rm heavy}Y_i^2 m_{3/2}^2 $.  The integrals
$I(m,n,l)$ are defined in the appendix.  The first set of entries correspond
to diagrams where $\tilde{X} \tilde{Y}$ insertions have been made in the
graphs.  The multiple listings of {\sl Graph 8-6} represent the additional contributions that arise when one or both of the 
supersymmetry-breaking vertices of a $m_{3/2} M \tilde{X}\tilde{Y}$ mass insertion is replaced
with supersymmetry-breaking gaugino mass vertex.  The {\sl 8-6 (non-local)}
and {\sl 10-1 (GMZ)} values enter alternately, depending on which
Lagrangian is used.  They are not to be added simultaneously toward
the total contribution.} 
\end{table}


\subsubsection{Finite Computation for $g'^4$}

Finally, a ``completely finite'' calculation, along the lines of that performed for the $(h^{*} h)^{2}$ case, is possible for the gauge
theory as well.  Although at first glance it appears that the theory
is unregulated, we can play the same game that we did in the Yukawa
theory, and imagine that we are regulating the theory through the use
of the Pauli-Villars regulators.  After the $X$ and $Y$ Pauli-Villars
regulator fields have been integrated out, a contribution proportional
to $Y_{X}^{2} Y_{\tau}^{2}$ arises, as shown in Equation~(\ref{eqn:gaugescalarmasshigh}).  
Integrating out the physical $X$ and  
$Y$ particles should precisely cancel this anomaly mediated
contribution.  This is the ultraviolet insensitivity.  This
calculation is outlined in an Appendix.  The result of the finite 
calculation of {\sl Graphs 8-1} through {\sl 8-8} yields the contribution: $\frac{1}{(4 \pi)^{4}}2
m_{3/2}^{2} g'^{4}(\mu) Y_{\tau}^{2} \sum_{\rm heavy}Y_i^2 $,
which precisely cancels the corresponding term in
Equation~(\ref{eqn:gaugescalarmasshigh}).

\section{Conclusion}
We have discussed how DRED can be used for calculations in anomaly
mediation.  Including operators proportional to $\epsilon$ is
absolutely essential, and failure to do so will yield incorrect
results.  For example, as we have shown, inclusion of the ${\cal
O}(\epsilon)$ operators is vital for recovering
ultraviolet-insensitivity.  We stress that inserting soft masses into
the Lagrangian by hand and calculating with the resulting piecemeal
Lagrangian will not give correct results.  The failsafe procedure is
to start with the bare Lagrangian given in Section \ref{sec:deriveL}
and compute from there.  The anomaly mediated soft terms seamlessly
emerge from these computations.

To demonstrate our DRED formalism, we have performed a diagrammatic
calculation to shed light on the anomaly mediated supersymmetry
breaking scenario.  In particular, we have shown explicitly how
threshold corrections keep the supersymmetry-breaking parameters on
anomaly-mediation trajectories.  This result is not a surprise,
considering the proof that already exists in the spurion calculus.
However, it is interesting to see exactly how the great multiplicity
of diagrams conspire to provide the necessary contributions and
cancellations.  Our calculation provides an explicit diagrammatic
check of the ultraviolet insensitivity.

Finally we mention that while the calculation in Section
\ref{sec:gauge} refers to an Abelian model, it would be relatively
straight-forward to extend this discussion to a non-Abelian gauge
theory.

\acknowledgements This was supported in part by the Director, Office
of Science, Office of High Energy and Nuclear Physics, Division of
High Energy Physics of the U.S. Department of Energy under Contract
DE-AC03-76SF00098 and in part by the National Science Foundation under
grant PHY-95-14797.  AP are EB are also supported by National Science
Foundation Graduate Fellowships. AP and HM would like to acknowledge
helpful conversations with Zoltan Ligetti, Aneesh Manohar, and Stephen Martin.  AP
would like additionally to acknowledge discussions with David Smith
and Yasunori Nomura.

\appendix
\section{Evaluation of the Integrals}\label{Appendix}

Several diagrams contain similar integrals.  In particular, it is 
useful to define:
\begin{equation}
I(m,n,l)=\frac{1}{(2 \pi)^{8}} \int{\frac{d^{4}p}{(p^{2}-M^{2})^m} 
\frac{d^{4}k}{(k^2-M^2)^n}\frac{1}{(k+p)^{2l}}},
\end{equation}
and it is convenient to regularize the integrals using dimensional
reduction. \footnote{Both integrals must be continued to $4- 2
\epsilon$ dimensions, even though in practice, one integral is
completely finite in the mass insertion formalism.  The reason is that
the ${\cal O}(\epsilon)$ terms in the finite integral can combine with
$\frac{1}{\epsilon}$ poles from the second integral to modify the
finite pieces in the result.}

After performing the integrals, $I(m,n,l)$ can be expressed entirely
in terms of Beta functions ($B$). In particular,
\begin{eqnarray}
& & I(m,n,l)=\frac{(-1)^{m+n+l+1}}{(4 \pi)^{4-\epsilon} 
(M^{2})^{n+m+l-4+\epsilon}} 
\frac{B(2,n+m-2)}{\Gamma(2-\epsilon) B(m,n)} \, \nonumber \\
& &\qquad \times B(n+l-2+\epsilon, m+l -2 +\epsilon) \, \nonumber \\
& &\qquad \times B(2-l-\epsilon, n+m+l-4+\epsilon).
\end{eqnarray}
This expression can then be Taylor expanded to order $\epsilon$, as
shown in Table VII.

\begin{table}[t] \label{table:seriesexpansion}
\begin{tabular}{l|l} 
Integral & Series Expansion \\ 
\\
$I(3,1,1)$ & \large{$ \frac{1}{M^{2}} (\frac{1}{2\epsilon} 
-\frac{\gamma}{2} + $ \ordere)} \\ 
$I(3,2,1)$ &  \large{$ \frac{1}{M^{4}} (-\frac{1}{4} +$ \ordere)} \\
$I(3,3,1)$ & \large{$ \frac{1}{M^{6}} (\frac{1}{12} +$ \ordere)} \\
$I(4,1,1)$ & \large{$ \frac{1}{M^{4}} (-\frac{1}{6 
\epsilon}-\frac{1}{12}+\frac{\gamma}{6}+$ \ordere)} \\
$I(4,2,1)$ & \large{$ \frac{1}{M^{6}} (\frac{1}{9}+$ \ordere)} \\
$I(5,1,1)$ & \large{$ \frac{1}{M^{6}} (\frac{1}{12 \epsilon} + 
\frac{1}{18} - \frac{\gamma}{12} +$ \ordere)}\\ 
\\
$I(3,2,0)$ & \large{$\frac{1}{M^{2}} (\frac{1}{2 \epsilon} - 
\frac{\gamma}{2}+$ \ordere)} \\
$I(3,3,0)$ & \large{$\frac{1}{M^{4}} (-\frac{1}{4} +$ \ordere)} \\
$I(4,1,0)$ & \large{$\frac{1}{M^{2}} (-\frac{1}{6 \epsilon} - 
\frac{1}{6} + \frac{\gamma}{6} +$ \ordere)} \\
$I(5,1,0)$ & \large{$\frac{1}{M^{4}} (\frac{1}{12 \epsilon} + 
\frac{1}{12} -\frac{\gamma}{12} +$ \ordere)} \\
\\
$G(3,3,1)$ & \large{$\frac{1}{M^{2}} (\frac{-1}{\epsilon} + \frac{5}{4} 
+ \gamma+ $ \ordere)} \\
& \\
$F(4,1,1)$ & \large{$\frac{1}{M^{2}} (\frac{2}{3 \epsilon} + 
\frac{1}{3} -\frac{2 \gamma}{3}+ $ \ordere)} \\
& \\
$F(5,1,1)$ & \large{$\frac{1}{M^{4}} (\frac{-1}{4 \epsilon} - 
\frac{5}{24} + \frac{\gamma}{4}+ $ \ordere)}
\end{tabular}
\caption{A list of series expansion useful in evaluation of $(h^{*} h)^{2}$ cancellations.  We neglect the Log($4\pi)$ and 
Log($M^{2}$) which cancel along with the Euler-$\gamma$'s.  There is 
also a common factor of $\frac{1}{(4 \pi)^{4}}$.}
\end{table}

We also also define the following integrals which are useful in the
evaluation of diagrams that include fermions:
\begin{equation}
F(m,n,l) \equiv \int{\frac{k \cdot (p-k)}{(p^{2})^{l}} 
\frac{d^{4}p}{((k-p)^2 -M^2)^{n}} \frac{d^{4}k}{(k^2 - M^2)^{m}}},
\end{equation}
and 
\begin{equation}
G(m,n,l) \equiv \int{\frac{k^{2} (k \cdot p)}{(p^{2})^{l}} 
\frac{d^{4}p}{((k+p)^2 -M^2)^{n}} \frac{d^{4}k}{(k^2 - M^2)^{m}}}.
\end{equation}
Using partial fraction decomposition, one can rewrite $F(m,n,l)$ and 
$G(m,n,l)$ in terms of $I(m,n,l)$.  In particular, we find that:
\begin{eqnarray}
F(m,n,l)&=& \frac{-1}{2} [I(m-1,n,l) +I(m,n-1,l) + \nonumber  \\
& &\phantom{\frac{-1}{2} [} I(m,n,l-1)] -M^{2} I(m,n,l), 
\end{eqnarray}
and 
\begin{eqnarray}
    \lefteqn{
    G(m,n,l) 
    = \frac{-1}{2} \times }\nonumber \\
    & &\bigl(I(m-1,n-1,l) - I(m-2,n,l) - I(m-1,n,l-1) 
    \nonumber \\
    & & + M^{2} [I(m,n-1,l)- I(m-1,n,l)-I(m,n,l-1)] \bigr).
    \nonumber \\
\end{eqnarray}

\section{Finite Calculations}\label{sec:finite}
Here we undertake the ``finite'' calculation of the $h^{4}$ correction
in the Yukawa model and a similar calculation in the gauge theory.
First consider the Yukawa theory.  Using Table II and Table VII, one
can write the threshold correction (excluding the $A$-term contribution)
as
\begin{eqnarray} 
    \mbox{Sum of Diagrams}=i(h^{*}(\mu)h(\mu))^{2}(M m_{3/2})^{2}
    \biggl((4\pi)^{-4} \frac{-11}{6M^2} \nonumber \\ 
    + 24 M^{4} I(5,1,1) + 30 M^{2} I(4,1,1)+ 6 I(3,1,1)\biggr).
\label{eqn:finiteform}
\end{eqnarray}
Here, we have only used the expressions in Table VII for those
integrals that are finite.  Since we are not working in DRED here,
there is no order $\epsilon$ contribution to these integrals.

Our remaining task is to calculate the combination $24 M^{4} I(5,1,1) 
+ 30 M^{2}
I(4,1,1)+ 6 I(3,1,1)$ without resorting to the regularization of any
integrals.  After a Wick rotation, we can
write this combination as:
\begin{eqnarray}
& & 6 \int \frac{d^{4} p \, d^{4}k}{(2 \pi)^{8}} 
\frac{1}{(k^{2}+M^{2})^{3}}\frac{1}{(k-p)^{2}+M^{2}}\frac{1}{p^{2}} 
\, \nonumber \\
& & \qquad \times \left( 1- \frac{5M^{2}}{k^{2}+M^{2}}+\frac{4 
M^{4}}{(k^{2}+M^{2})^{2}}\right).
\end{eqnarray}
This, in turn, can be written as:
\begin{equation}
-6 \int \frac{d^{4} p \, d^{4}k}{(2 \pi)^{8}} 
\frac{1}{(k-p)^{2}+M^{2}}\frac{1}{p^{2}} 
\frac{1}{k^{2}}\frac{\partial}{\partial k^{2}} 
\frac{(k^{2})^{3}}{(k^{2}+M^{2})^{4}}.
\end{equation}
Now this integral can be done by first doing the $k^{2}$ integral by 
parts.  The surface term vanishes, leaving 
\begin{equation}
-3 \int \frac{d^{4} p \; d^{4}k}{(2 \pi)^{8}} \frac{2k^{4} -2k^{2}(p 
\cdot k)}{\bigl((k-p)^{2}+M^{2} \bigr)\, p^{2}}
\frac{1}{(k^{2}+M^{2})^{4}}.
\end{equation}
From this point, standard Feynman parameter techniques can be 
employed, yielding the result:
\begin{eqnarray}
& &24 M^{4} I(5,1,1) + 30 M^{2} I(4,1,1)+ 6 I(3,1,1) \nonumber \\
& & \qquad =(4\pi)^{-4} \frac{-7}{6M^2}.
\end{eqnarray} 
Combining this with Equation~(\ref{eqn:finiteform}) yields
\begin{equation}
\mbox{Sum of Diagrams}=-3i(h^{*}(\mu) h(\mu))^{2}\frac{m_{3/2}^{2}}{(4\pi)^{4}} .
\end{equation}
As discussed in the text, this combines with a contribution
$+6i(h^{*}(\mu) h(\mu))^{2}\frac{1}{(4\pi)^{4}}m_{3/2}^{2}$ from the $A$-term diagram to
yield the correct threshold correction for the scalar mass.

Incidentally, the calculation of the same integrals in dimensional 
regularization will yield 
$\frac{-19}{6M^{2}}(4\pi)^{-4}$.  The 
difference results from the fact that the $d^{4}k$ becomes a 
$d^{4-2\epsilon}k$ and the integration by parts picks up an extra 
piece.
 
Now consider the gauge theory.  Again, the game will be to keep all
integrals well defined without ever continuing to $4-2 \epsilon$
dimensions.  Since we stay in 4 dimensions, the evanescent operators
do not arise, and we need only consider {\sl Graphs 8-1, 8-3, 8-6},
and {\sl Graph 8-7}.  The key is combine these graphs first, avoiding
any divergent (ill-defined)integrals.

{\sl Graph 8-3} can be written as:
\begin{eqnarray}
& &\mbox{\sl Graph 8-3}= m_{3/2}^{2} M^{2} {g^{\prime}}^{4}(\mu)
Y_{\tau}^{2}\sum_{\rm heavy} Y_i^2\frac{1}{(4 \pi)^{2}}\times \nonumber \\
& & \qquad \int
\frac{d ^{4} p}{(2 \pi)^{4} p^{2}} \int \frac{ z (2z-1) \, dz}{(M^{2}
-p^{2} z (1-z))^{2}}.
\label{eqn:finitesettingsun}
\end{eqnarray}
By itself, this integral would be divergent at the endpoints of the
Feynman parameter integral.  So we must combine this expression with
expressions for the remaining graphs before evaluation.  {\sl Graph
8-6} can be written:
\begin{eqnarray}\label{eqn:finitegaugino}
& &\mbox{\sl Graph 8-6}= 4 m_{3/2}^{2} M^{2} {g^{\prime}}^{4}(\mu)
Y_{\tau}^{2}\sum_{\rm heavy}Y_i^2\frac{1}{(4 \pi)^{2}}\times \nonumber \\
& & \qquad \int
\frac{d ^{4} p}{(2 \pi)^{4} p^{2}} \int \frac{ z^{3} \, dz}{(M^{2}
-p^{2} z (1-z))^{2}}.
\end{eqnarray}
Finally, we write the sum of {\sl Graphs 8-1} and {\sl 8-7} using
the vacuum polarization operator:
\begin{equation}
\label{eqn:finiteGI}
\mbox{\sl Graph 8-1+ Graph 8-7}=-3 i {g^{\prime}}^{2}(\mu) \int \frac 
{\Pi(p^{2}) \, d^{4} p}{(2 \pi)^{4} \, p^{2}},
\end{equation}
where $\Pi(p^{2})$ is the vacuum polarization operator to ${\cal
O}(m_{3/2}^{2})$ in four dimensions.  In the mass insertion formalism,
 it is given by the expression:
\begin{equation}
\label{eqn:finitevacpol}
\Pi(p^{2}) \equiv \frac{i m_{3/2}^{2} M^{2} (Y_{X_{1}}^{2}
 +Y_{X_{2}}^{2}) {g^{\prime}}^2(\mu)}{(4 \pi)^{2}} \int \frac{ (1-2z)^{2}
 \, dz}{(M^{2} -p^{2} z (1-z))^{2}}, 
\end{equation} 
which can be seen by taking the $\epsilon \to 0$ limit in
Equation~(\ref{eqn:vacuumpolarMI}).  

Utilizing Equations~(\ref{eqn:finitesettingsun}), 
(\ref{eqn:finitegaugino}), (\ref{eqn:finiteGI}), 
and (\ref{eqn:finitevacpol}), we can write
the sum of diagrams as
\begin{eqnarray}
&&\mbox{{\sl Graph 8-1} + {\sl 8-3} + {\sl 8-6} + {\sl 8-7}}=-4 i\frac{1}{(4\pi)^{4}} 
m_{3/2}^{2} M^{2} \nonumber \\
&& 
\qquad \times {g^{\prime}}^{4}(\mu) Y_{\tau}^{2}\sum_{\rm heavy}Y_i^2
\int \int\limits_{0}^{1} \frac{z (1-z)^{2} \, dz
\, d^{4} p}{p^{2} (M^{2} -p^{2} z (1-z))^{2}}.
\end{eqnarray}
This integral is completely finite so no regulator is needed.  The
integral yields a contribution to the scalar mass
\begin{eqnarray}
& &-\frac{1}{i} (\mbox{{\sl Graph 8-1} + {\sl 8-3} + {\sl 8-6} + 
{\sl 8-7}})\nonumber \\
& &\qquad = \frac{1}{(4 \pi)^{4}}2 m_{3/2}^{2} M^{2} {g^{\prime}}^{4}(\mu)
Y_{\tau}^{2}\sum_{\rm heavy}Y_i^2.
\end{eqnarray}
This precisely corrects Equation~(\ref{eqn:gaugescalarmasshigh}) to be
Equation~(\ref{eqn:gaugescalarmasslow}), demonstrating the ultraviolet
insensitivity.

\section{Calculation of Wave-Function Renormalization in 
Gauge Theory} 

To offer an alternative to direct computation, and to avoid the
niceties of a supersymmetric (sans Wess-Zumino gauge) calculation of
$Z_{\tau}$, we can use renormalization group principles to determine
$\tilde m_{\tau}^2$ in terms of bare couplings.  Working above
threshold, the generic structure of two-loop diagrams tells us that
$Z_{\tau}$ must be of the form
\begin{equation}\label{eqn:ztauform}
Z_{\tau} = 1 + A \frac{Y_{\tau}^2
g'^2_0}{(4\pi)^2(\mu^2)^{\epsilon}}\frac{1}{\epsilon} + B
\frac{Y_{\tau}^2
g'^4_0}{(4\pi)^4(\mu^2)^{2\epsilon}}\frac{1}{\epsilon^2}.
\end{equation}
Then
\begin{eqnarray}
&&\gamma_{\tau} \equiv -\frac{1}{2}\mu \frac{d}{d\mu} \log
Z_{\tau}\nonumber\\
&&  = A \frac{Y_{\tau}^2
g'^2_0}{(4\pi)^2(\mu^2)^{\epsilon}} + 2 B \frac{Y_{\tau}^2 g'^4_0}
{(4\pi)^4(\mu^2)^{2\epsilon}}\frac{1}{\epsilon} - A^2 \frac{Y_{\tau}^4
g'^4_0} {(4\pi)^4(\mu^2)^{2\epsilon}}
\frac{1}{\epsilon}. \nonumber \\
\label{eqn:gammatauform}
\end{eqnarray}
The poles in $\gamma_{\tau}$ are lower order than the poles in
$Z_{\tau}$ because $\mu$ derivatives hitting terms like
$(\mu^2)^{\epsilon}$ bring down factors of $\epsilon$.

We know that in the $\epsilon\to 0$ limit, the expression for
$\gamma_{\tau}$ must agree with the expression in terms of the
renormalized coupling to one-loop order.  Comparing with
Equation~(\ref{eqn:gammataugauge}) fixes $A = -2$, so that
\begin{eqnarray}
\label{eqn:gammataufixedA}
\gamma_{\tau} &=& -2 \frac{Y_{\tau}^2
g'^2_0}{(4\pi)^2(\mu^2)^{\epsilon}} \nonumber \\
& &+ 2 B \frac{Y_{\tau}^2 g'^4_0}
{(4\pi)^4(\mu^2)^{2\epsilon}}\frac{1}{\epsilon} - 4 \frac{Y_{\tau}^4
g'^4_0} {(4\pi)^4(\mu^2)^{2\epsilon}}\frac{1}{\epsilon}.
\end{eqnarray}

Now we work to fix $B$.  We can do this by utilizing two pieces of
information: the known expression for the running of the gauge
coupling and the finiteness of $\gamma_{\tau}$.  To proceed we first
write the bare coupling in terms of the renormalized coupling.  They
are equal at one loop, and at higher order we include an arbitrary
parameter $C$ to be completely general.  We define the renormalized
coupling as:
\begin{equation}
g'^2(\mu) + C g'^4(\mu)
\equiv g'^2_0  - B \frac{ g'^4_0}
{(4\pi)^2(\mu^2)^{\epsilon}}\frac{1}{\epsilon} +2 \frac{Y_{\tau}^2
g'^4_0} {(4\pi)^2(\mu^2)^{\epsilon}}\frac{1}{\epsilon}.\label{eqn:defnC}
\end{equation}
This form is convenient because it allows us to rewrite the
$\gamma_{\tau}$ of Equation~(\ref{eqn:gammataufixedA}) simply in terms
of renormalized couplings,
\begin{equation}
\gamma_{\tau} = -2 \frac{Y_{\tau}^2
g'^2(\mu)}{(4\pi)^2(\mu^2)^{\epsilon}}+ C \frac{Y_{\tau}^2
g'^4(\mu)}{(4\pi)^4 (\mu^2)^{\epsilon}}.
\end{equation}
Now we make the critical observation that $\gamma_{\tau}$ is an 
observable
quantity, and so it must be finite in the $\epsilon\to 0$ limit when
expressed in terms of the renormalized coupling.  In other words, $C$
is finite.  Now we can determine $B$ by comparing our definition in
Equation~(\ref{eqn:defnC}) with the known running of the gauge coupling
constant:
\begin{equation}
g'^2(\mu) = g'^2_0 - \frac{g'^4_0 (Y_{\tau}^2 +
\sum_{\rm heavy}Y_i^2)}{(4\pi)^2(\mu^2)^{\epsilon}}\frac{1}{\epsilon}.
\end{equation}
Inserting this known expression for $g'^2(\mu)$ into
Equation~(\ref{eqn:defnC}) and keeping up to $O(g'^4)$, we find the
condition
\begin{eqnarray}
\lefteqn{
g'^2_0 - \frac{g'^4_0 (Y_{\tau}^2 +
\sum_{\rm heavy}Y_i^2)}{(4\pi)^2(\mu^2)^{\epsilon}}\frac{1}{\epsilon} + C
g'^4_0} \nonumber \\ 
& &= g'^2_0 - B \frac{ g'^4_0}
{(4\pi)^2(\mu^2)^{\epsilon}}\frac{1}{\epsilon} +2 \frac{Y_{\tau}^2
g'^4_0} {(4\pi)^2(\mu^2)^{\epsilon}}\frac{1}{\epsilon}.
\end{eqnarray}
Equivalently, 
\begin{equation}
B = (Y_{\tau}^2 + \sum_{\rm heavy}Y_i^2) + 2 Y_{\tau}^2 - C \epsilon
(4\pi)^2(\mu^2)^{\epsilon}. 
\end{equation}
But since $C$ is finite and comes multiplied by $\epsilon$, it makes
at most a finite $O(g'^4)$ contribution to $\gamma_{\tau}$, which is
next-to-leading order in Equation~(\ref{eqn:gammatauform}).  We have
consistently been neglecting such terms.  In short,
\begin{equation}
B = (Y_{\tau}^2 + \sum_{\rm heavy}Y_i^2) + 2 Y_{\tau}^2. 
\end{equation}

We have determined that
\begin{equation}
\gamma_{\tau} = -2 \frac{Y_{\tau}^2 g'^2_0}{(4\pi)^2(\mu^2)^{\epsilon}}
+ 2 \frac{ g'^4_0 Y_{\tau}^2 (Y_{\tau}^2 + \sum_{\rm heavy}Y_i^2)}
{(4\pi)^4(\mu^2)^{2\epsilon}}\frac{1}{\epsilon},
\end{equation}
and differentiation yields the scalar mass in terms of bare couplings:
\begin{eqnarray}
\label{eqn:gaugescalarmasshighbareapp}
\tilde{m}^{2}_{\tau} &=& \frac{m_{3/2}^2}{(4\pi)^2} (2 \frac{\epsilon
Y_{\tau}^{2} {g'}^{2}_0}{ (\mu^2)^{\epsilon}} -4\frac{{g'}^{4}_0
Y_{\tau}^2 (Y_{\tau}^2 + \sum_{\rm heavy} Y_{i}^2)}{(4\pi)^2
(\mu^2)^{2\epsilon}}) \\ && {(\mbox{Above Threshold, Bare Couplings
})}.\nonumber
\end{eqnarray}
Below threshold, $\mu\ll M$, and the analysis is similar.  In
$\gamma_{\tau}$ we make the replacement $Y_{i}^2 
(\mu^2)^{-\epsilon}\to
Y_{i}^2 (M^2)^{-\epsilon}$ for the heavy particles, because their
contributions to loop integrals are cut off at $M$:  We then find
\begin{eqnarray}
\gamma_{\tau} &\to& -2 \frac{Y_{\tau}^2
g'^2_0}{(4\pi)^2(\mu^2)^{\epsilon}} \nonumber \\
& &+ 2 \frac{ g'^4_0 Y_{\tau}^4 }
{(4\pi)^4(\mu^2)^{2\epsilon}}\frac{1}{\epsilon} + 2 \frac{ g'^4_0
Y_{\tau}^2 \sum_{\rm heavy}Y_i^2 }
{(4\pi)^4(\mu^2)^{\epsilon}(M^2)^{\epsilon}}\frac{1}{\epsilon}.
\end{eqnarray}
Clearly then
\begin{eqnarray}
\label{eqn:gaugescalarmasslowbareapp}
\tilde{m}^{2}_{\tau} &=& \frac{m_{3/2}^2}{(4\pi)^2} (2 \frac{\epsilon
Y_{\tau}^{2} {g'}^{2}_0}{ (\mu^2)^{\epsilon}} 
\nonumber \\
& &-4\frac{{g'}^{4}_0
Y_{\tau}^4}{(4\pi)^2
(\mu^2)^{2\epsilon}} -2\frac{{g'}^{4}_0 Y_{\tau}^2 \sum_{\rm heavy}
Y_{i}^2}{(4\pi)^2 (\mu^2)^{\epsilon}(M^2)^{\epsilon}}) \\ & &
{(\mbox{Below Threshold,
Bare Couplings })}.\nonumber
\end{eqnarray}
%

\end{document}